\documentclass[sigconf, nonacm]{acmart}

\renewcommand\footnotetextcopyrightpermission[1]{} 

\usepackage{listings,amsfonts}
\usepackage{amsmath,xcolor,pifont}
\usepackage{footnote}
\makesavenoteenv{table}
\usepackage{url}
\usepackage{framed}
\usepackage{balance}
\usepackage{xspace}
\usepackage{bbding}
\usepackage{multirow}
\usepackage{hyperref,epsfig,endnotes}
\usepackage{xcolor}
\usepackage{booktabs}
\usepackage{array}
\usepackage{algorithm}
\usepackage{paralist}
\usepackage{multirow,makecell}
\usepackage{enumerate}
\usepackage{diagbox}
\usepackage{fancynum}
\usepackage{subfig}

\usepackage{amssymb}
\usepackage{pifont}
\newcommand{\cmark}{\ding{51}}%
\newcommand{\xmark}{\ding{55}}%

\newcommand{\tabincell}[2]{\begin{tabular}{@{}#1@{}}#2\end{tabular}}

\hypersetup{draft}

\newcommand{\Rplus}{\protect\hspace{-.1em}\protect\raisebox{.35ex}{\smaller{\smaller\textbf{+}}}}
\newcommand{\Cpp}{\mbox{C\Rplus\Rplus}\xspace}

\renewcommand\footnotetextcopyrightpermission[1]{} 

\newcommand{\latinlocution}[1]{\textit{#1}}
\newcommand{\eg}{\latinlocution{e.g.,}\xspace}
\newcommand{\ie}{\latinlocution{i.e.,}\xspace}

\newcolumntype{L}[1]{>{\raggedright\let\newline\\\arraybackslash\hspace{0pt}}m{#1}}
\newcolumntype{C}[1]{>{\centering\let\newline\\\arraybackslash\hspace{0pt}}m{#1}}
\newcolumntype{R}[1]{>{\raggedleft\let\newline\\\arraybackslash\hspace{0pt}}m{#1}}

\usepackage{adjustbox}
\newcolumntype{R}[2]{%
    >{\adjustbox{angle=#1,lap=\width-(#2)}\bgroup}%
    l%
    <{\egroup}%
}

\renewcommand{\texttt}[1]{%
 \begingroup
 \ttfamily
 \begingroup\lccode`~=`/\lowercase{\endgroup\def~}{/\discretionary{}{}{}}%
 \begingroup\lccode`~=`[\lowercase{\endgroup\def~}{[\discretionary{}{}{}}%
 \begingroup\lccode`~=`.\lowercase{\endgroup\def~}{.\discretionary{}{}{}}%
 \catcode`/=\active\catcode`[=\active\catcode`.=\active
 \scantokens{#1\noexpand}%

 \endgroup
 }

\begin{document}

\renewcommand{\figurename}{Fig.}
\title{Characterizing EOSIO Blockchain}

\author{Yuheng Huang}
\affiliation{%
  \institution{Beijing University of Posts and Telecommunications}
}

\author{Haoyu Wang}
\authornote{Corresponding Authors: Haoyu Wang (haoyuwang@bupt.edu.cn), and Lei Wu (lei\_wu@zju.edu.cn)}
\affiliation{%
  \institution{Beijing University of Posts and Telecommunications}
}

\author{Lei Wu}
\authornotemark[1]
\affiliation{%
  \institution{Zhejiang University}
}

\author{Gareth Tyson}
\affiliation{%
  \institution{Queen Mary University of London}
}

\author{Xiapu Luo}
\affiliation{%
  \institution{The Hong Kong Polytechnic University}
}

\author{Run Zhang}
\affiliation{%
  \institution{Beijing University of Posts and Telecommunications}
}

\author{Xuanzhe Liu}
\affiliation{%
  \institution{Peking University}
}

\author{Gang Huang}
\affiliation{%
  \institution{Peking University}
}

\author{Xuxian Jiang}
\affiliation{%
  \institution{PeckShield, Inc.}
}

\renewcommand{\shortauthors}{Huang and Wang, et al.}

\begin{abstract}

EOSIO has become one of the most popular blockchain platforms since its mainnet launch in June 2018.
In contrast to the traditional PoW-based systems (e.g., Bitcoin and Ethereum), which are limited by low throughput, EOSIO is the first high throughput Delegated Proof of Stake system that has been widely adopted by many applications.  
Although EOSIO has millions of accounts and billions of transactions, little is known about its ecosystem, especially related to security and fraud. 
In this paper, we perform a large-scale measurement study of the EOSIO blockchain and its associated DApps. We gather a large-scale dataset of EOSIO and characterize activities including money transfers, account creation and contract invocation.
Using our insights, we then develop techniques to automatically detect bots and fraudulent activity. 
We discover thousands of bot accounts (over 30\% of the accounts in the platform) and a number of real-world attacks (301 attack accounts). By the time of our study, 80 attack accounts we identified have been confirmed by DApp teams, causing 828,824 EOS tokens losses (roughly 2.6 million US\$) in total.

\end{abstract}

\maketitle

\section{Introduction}
Blockchain technologies, such as Bitcoin~\cite{bitcoin}, have experienced much hype in recent years. Blockchain technologies have been proposed for applications in many areas, including financial and logistical systems. For example, the Internet giant Facebook recently announced their plans for a cryptocurrency~\cite{libra}. 
Blockchain technologies have found particular popularity with decentralized application (DApp) developers, most notably for creating smart contracts.
These consist of a decentralised protocol which is capable of digitally negotiating an agreement in a cryptographically secure manner. As the most widely used blockchain system after Bitcoin, Ethereum~\cite{ethereum} offers support for executing smart contracts, yet it suffers from poor performance (as with Bitcoin) due to their reliance on Proof-of-Work (PoW) consensus protocols. 

This has motivated researchers to propose new blockchain approaches that employ more efficient consensus mechanisms. 
One particularly prominent example is \emph{EOSIO}, the largest Initial Coin Offering (ICO) project to date (over \$4 billion). EOSIO adopts a Delegated Proof-of-Stake (DPoS) consensus protocol. This allows EOSIO to achieve far higher performance throughput, i.e., up to $8,000$ Transactions Per Second (TPS) within a single thread, and unlimited for multiple-threaded cases~\cite{EOSIOTPS}. As a result, EOSIO has grown rapidly and successfully surpassed Ethereum in DApp transactions just three months after its launch (in June 2018).
For example, a recent report demonstrated that the average amount of EOS (the EOSIO currency) traded in 24 hours has achieved $57$ million (with a peak exceeding $80$ million)~\cite{cryptonomist}. As a comparison, Bitcoin has an average of 825 thousand transactions and Ethereum has an average of 717 thousand transactions.

Consequently, EOSIO has attracted significant attention from both industry and research communities alike.
Criticisms, however, have started to emerge, accusing EOSIO of suffering from \textit{superficial prosperity}, \ie that it has a large number of transactions, yet the majority of users are inactive~\cite{8btc,anchain}. 
Furthermore, recent years have witnessed attacks against EOSIO, exploiting vulnerabilities in DApps. This has resulted in millions of dollars lost~\cite{eos-attack1, eos-attack2, eos-attack3}. 
Despite these anecdotes~\cite{lee2019spent, quan2019evulhunter}, \emph{we still lack a comprehensive understanding of EOSIO's operation in the wild,} especially the \textit{severity} of problems that EOSIO faces.

To rectify this, we present a detailed study of the EOSIO ecosystem at scale, longitudinally and across various dimensions. To this end, we first gather a large-scale dataset containing both on-chain data of EOSIO and off-chain data related to DApps and attacks (\textbf{Section~\ref{sec:dataset}}). Our dataset consists of over 3 billion transactions, over 1 million EOSIO accounts, thousands of bots and a number of attack reports. Based on the collected dataset, we then perform an explorative study to characterize the activities on EOSIO (\textbf{Section~\ref{sec:general}}), including money transfer, account creation, and contract invocation. 
Following this, we identify bot-like accounts and fraudulent activities by mining the relations and behavioral similarities among millions of accounts (\textbf{Section~\ref{sec:groupcontrol}}). 
We further investigate their incentives and purposes.
Finally, we characterize security issues in EOSIO, including permission misuse issues and attacks (\textbf{Section~\ref{sec:security}}). 

To the best of our knowledge, this is the first comprehensive study of the EOSIO blockchain at \emph{scale}, \emph{longitudinally},  and \emph{across various dimensions}. We have revealed a range of serious misbehaviors on the EOSIO blockchain. Among many interesting results and observations, the following are prominent:

\begin{itemize}
    \item \textbf{The overall ecosystems follows the Pareto principle.}
    Although the overall ecosystem shows a growing volume of transactions (over 1 billion transfers), EOSIO is dominated by a small percentage of accounts. The top $0.47\%$ of accounts constitute $90\%$ of the total transaction volume. Exchanges and gambling DApps dominate the transactions. \textbf{Over 32\% of the accounts are ``silent''}: they have never actively initiated any transaction with other accounts. 
    \item \textbf{Bot-like accounts are prevalent}. We flag over 30.75\% of the accounts (381,008) as bot-like, with over 192 million transactions, and 640 million EOS transferred. These bots are mainly used for malicious and fraudulent purposes including Bonus Hunting, Clicking Fraud, etc.
    
    \item \textbf{Permission misuse issues are overlooked by users.} 
    We identify permission misuse issues of 5,541 accounts, i.e., granting their ``eosio.code'' permissions to other accounts, which could cause serious security issues (e.g., the accounts with the granted permission can stealthily transfer users' EOS tokens even without their attentions). 

    \item \textbf{EOSIO suffers from a number of serious attacks.} 
    We identify over 301 suspicious attack accounts, causing over 1.5 million EOS losses. We have reported them to the DApp teams (developers): 80 of the attacks (with 828,824 EOS losses) have been confirmed by the time of our study. We further assist the DApp teams in tracing the losses. 
    
\end{itemize}

We develop core methodologies to trace attacks and fraud, as well as deriving key insights into EOSIO. Our efforts contribute developer awareness, inform the activities of the research community and regulators, and promote better operational practices across blockchains.

\section{Background}
\label{sec:background}
We start by briefly presenting  key concepts. Note that this is intended as an overview; we refer readers to~\cite{eos-doc} for full details.

\subsection{Overview of EOSIO}

EOSIO is a decentralized enterprise system that executes industrial-scale DApps~\cite{eos} --- software which relies on the EOSIO blockchain to cryptographically record \emph{transactions}. 
The most common transaction is transferring the EOSIO currency token, named \textit{EOS}.
In contrast to Bitcoin or Etherum, EOSIO is a DPoS-based system, which can scale to millions of transactions per second, making EOSIO an attractive option for new DApp developers.

There are four key concepts to understand within EOSIO. If an entity wishes to interact with the EOSIO blockchain, it must first create an \emph{account}. This unique identity can then invoke a \emph{smart contract} through a \emph{transaction}, which consists of one more \emph{actions} to perform. For example, an action might be transferring an EOS token from one account holder to another.
To enable this process, accounts wishing to invoke a contract must first delegate appropriate \emph{permissions}, granting it the privileges to act on its behalf. The rest of this section describes in detail these four concepts.

\subsection{EOSIO's Smart Contract and Transactions}
EOSIO adopts \Cpp as the official language for DApp developers to write smart contracts. In particular, the contract source code is first compiled down to WebAssembly (aka WASM) bytecode. Upon invocation, the bytecode will then be executed in EOSIO's WASM VM, resulting in transactions recorded on the blockchain, \eg transferring EOS. Note that an account can be associated with no more than one smart contract.

As the basic element of communication between smart contracts, an \textit{action} is a base32 encoded 64-bit integer which can be used to represent a single operation. There are two types of actions: \emph{external action},  when the user calls an action directly from the outside; and an \emph{inline action}, where an inline action refers to a call to another action in a smart contract (same or external). 

Each transaction can consist of one or more actions. In EOSIO, there are two ways to send actions for communication: an \textit{inline} action that performs an operation within the same transaction as the original action, and \textit{deferred} action that might be scheduled to perform operation in a future (deferred) transaction. Inline actions are guaranteed to execute synchronously, while deferred actions will be executed asynchronously if being scheduled. The transaction will be rolled back if an inline action fails or raises exception.

\subsection{EOSIO's Account Management}
\label{subsec:account}

In EOSIO, accounts are the entities that can execute transactions. An EOSIO account is a human-readable name (up to 12 characters) recorded on the blockchain. In practice, accounts are authorization structures that can define senders and receivers of contracts. Accounts can also grant permissions to contracts and be configured to provide individual or group access to transfer/push any valid transactions to the blockchain.

Note that an EOSIO account is different from (and more complicated than) that of Ethereum. Most notably, accounts are hierarchical and can only be created by an existing account, thereby creating a tree structure.
This means the resources required to create new accounts must be allocated by existing accounts (except the first privileged account named \textit{eosio} --- the root of the tree --- created by the blockchain system when the mainnet was launched). 
This inevitably consumes system resource (RAM) and therefore the account creation of EOSIO is not free.\footnote{More precisely, one has to buy RAM to store the account data~\cite{eos-dev-glossary}.}

Permissions associated with an EOSIO account are used to authorize actions and transactions to other accounts~\cite{eos-dev-accounts-permissions}. 
Specifically, the account can assign public/private keys to specific actions, and a particular key pair will only be able to execute the corresponding action. By default, an EOSIO account is attached to two public keys: the \textit{owner} key (which specifies the ownership of the account) and the \textit{active} key (which grants access to activities with the account). These two keys authorize two native named permissions: the \textit{owner} and \textit{active} permission, to manage accounts. Apart from the native permissions, EOSIO also allows customized named permissions for advanced account management.

\subsection{Example}

\begin{figure}[!htbp]
    \centering
    \includegraphics[width=0.48\textwidth]{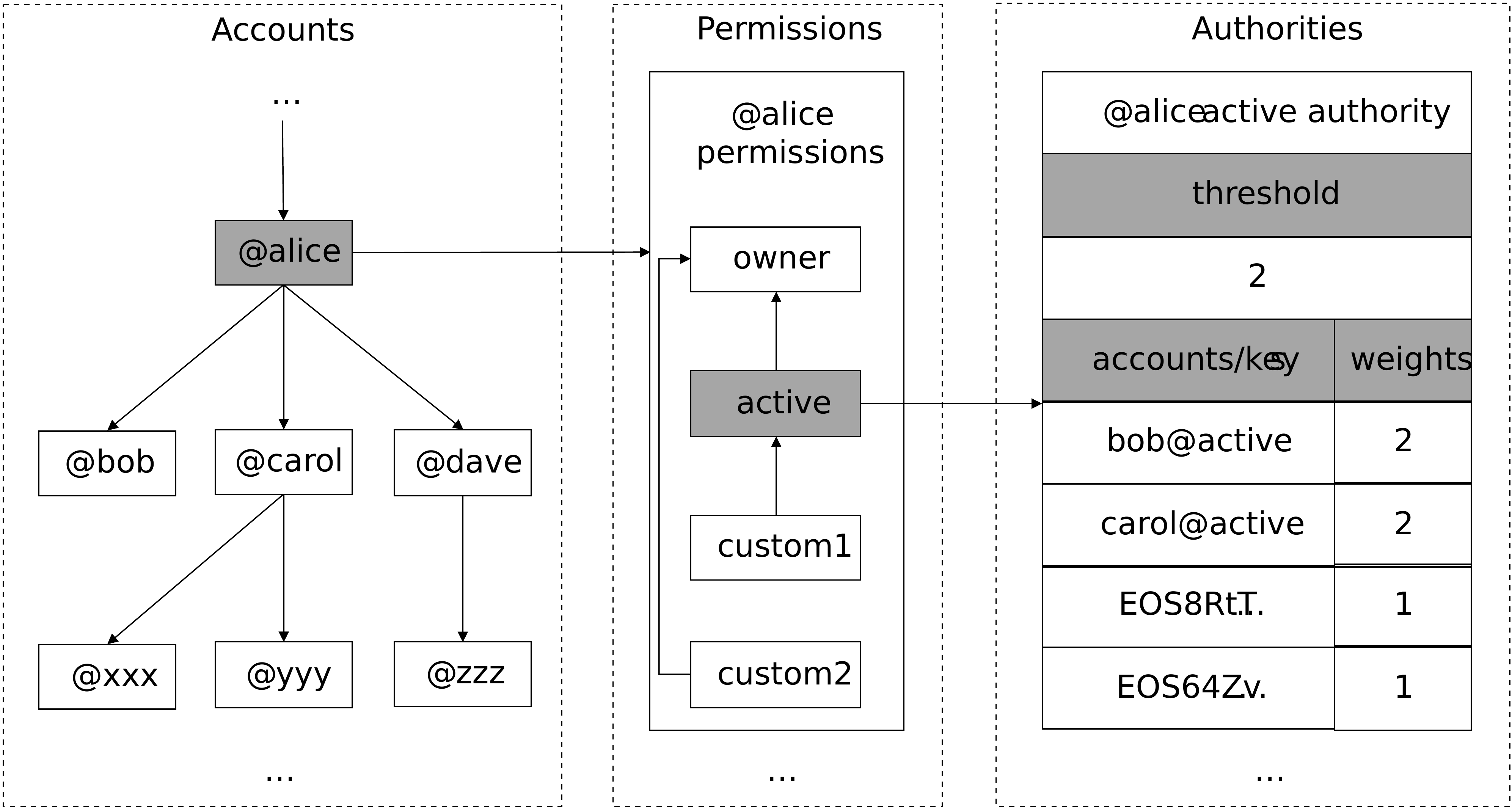}
    \caption{An Example of EOSIO Accounts.}
    \label{fig:eos_account}
\end{figure}

Fig.~\ref{fig:eos_account} gives an example to demonstrate the relationship between accounts, associated permissions and authorities.
These account permissions can be delegated to actions, such that transactions can be performed on their behalf. 
The left most part of the figure shows the hierarchical structure (i.e., tree) of the accounts. The account named \textit{alice} (marked as grey) is the root of the tree. 
Alice has created three other accounts: bob, carol and dave.
The central part of the figure shows that \textit{alice} has 4 permissions, including two native permissions (\textit{owner} and \textit{active}) and two customized permission (\textit{custom1} and \textit{custom2}) respectively. 
The right most part of the figure gives the authority table of the grey marked permission \textit{active}. 
This cyptographically states which permissions have been allocated to the children accounts of alice. 
Note that there exists a threshold that must be reached to authorize the execution of the action. In addition, in order to be executed by or on behalf of \textit{alice}, a weight threshold of 2 must be reached as well~\cite{eos-dev-accounts-permissions}. The permissions and authorities will be used in Section~\ref{sec:permission}.

\section{Study Design \& Data Collection}
\label{sec:dataset}

We seek to focus on the following three research questions (RQs):

\begin{itemize}
    \item[RQ1] \textbf{What are the characteristics of accounts and their transaction behaviors in the EOSIO ecosystem? } No previous work has characterized the EOSIO ecosystem, including the health of the platform and the various crypto-assets within. As previous work has investigated Ethereum~\cite{TingINFOCOM18}, we wish to compare them and understand the difference between these two blockchain platforms.
    
    \item[RQ2] \textbf{How severe is the presence of bot activities in the ecosystem?} 
    As many applications experience a range of bot activities~\cite{InternetBot}, we wish to investigate whether blockchains are also inundated with bot activities. Although millions of transactions are emerging on the EOSIO platform, it is unknown \emph{how many of them are manipulated by bots and how many accounts/activities are fake}.
    
    \item[RQ3] \textbf{Can we identify real-world security issues by analyzing the accounts and transactions we collect?} 
    As one of the most popular platforms for DApps, EOSIO has always been the target for hackers. A number of reports have already revealed attacks, leading to millions of dollars lost. Thus, it is interesting to explore whether we can identify real-world attacks and build an early warning system.
    
\end{itemize}

\subsection{Data Collection}

To provide a comprehensive analysis of EOSIO, we first seek to harvest both \textit{on-chain data} and \textit{off-chain information} (cf. Table~\ref{table:collceted_data}):

\begin{enumerate}[1)]
\item \textbf{On-chain data of EOSIO blockchain}:
\begin{itemize}
    \item \textbf{Transaction Records.} To enforce a fine-grained analysis, we use \emph{actions} to measure the transaction activities because actions are the basic unit that constitute a transaction in EOSIO, as mentioned in Section~\ref{sec:background}. However, due to the volume of on-chain data, it is not feasible to fetch them either by querying public API endpoints or by crawling from the blockchain explorer. To solve the problem, we have built a customized EOSIO client, which can be used to synchronize with the mainnet in an efficient way.
    Note that we also collected all the notifications on the EOSIO blockchain, which could be used to facilitate the detection of specific kinds of attacks (e.g., Fake EOS Transfer attack, cf. Section~\ref{sec:security}).

    \item \textbf{Account Information.} 
    An EOSIO account has many attributes, including creation information (e.g., creator and creation time), owned system resources, assigned keys and permissions. The creation information can be traced by crawling blockchain explorers (we take advantage of EOSPark). Other information can be queried through public API endpoints~\cite{APIEndpoint}. Many engineering efforts were required to overcome the heterogeneous structures of different explorers and APIs. To this end, we have implemented a generic collector that is capable of crawling and querying necessary account information. 
\end{itemize}

\vspace{0.1in}
\item \textbf{Off-chain data related to DApps, Bots and Attacks}: 
\begin{itemize}
    \item \textbf{DApp Information.} As accounts on the EOSIO all have human-readable names, finding whether an account belongs to a DApp is more straightforward than Ethereum. By crawling data from websites including DAppTotal~\cite{dapptotal} and DAppReview~\cite{dappreview}, we have annotated accounts with their DApp. To the best of our knowledge, we have collected the most complete list of Dapp accounts available.

    \item \textbf{Bot Accounts.} Blockchain bots have been revealed by DApp teams and blockchain security companies~\cite{anchain, blockchainBot}. We have collected 63,956 bot-like accounts, which will be used as our ground-truth in bot accounts identification (cf. Section~\ref{sec:groupcontrol}). 
    
    \item \textbf{Attack Information.} A number of attacks on EOSIO have already been observed in the wild. To measure the severity, we collect existing attack information, including date, participants (victims and attackers) and damage, by monitoring and collecting security news and blogs from well-known blockchain security companies. Based on this ground-truth dataset, we implement a monitoring system to perform attack detection and forensics (cf. Section~\ref{sec:security}). 
\end{itemize}

\end{enumerate}

In summary, we have collected transactions from $2018.06.09$ to $2019.05.31$, with over 3 billion actions in total. We also collected information for all 1,239,030 accounts (including $2,482,192$ keys).
Moreover, we have collected
$1,513$ DApp accounts, 63,956 bot-like accounts and 40 attack events (including 37 attack accounts).

\begin{table}[t]
\centering
\caption{Overview of our dataset (on-chain and off-chain).}
\vspace{-0.1in}
\label{table:collceted_data}
\setlength{\tabcolsep}{7mm}{
\resizebox{0.8\linewidth}{!}{
\begin{tabular}{r|r}
\hline
\textbf{Category} & \textbf{Amount}  \\
\hline \hline
Action trace & $3,208,168,339$ actions\\
Account information & $1,239,030$ accounts\\
\hline
DApp accounts & $1,513$ accounts\\
Bot accounts &  $63,956$ accounts \\
Attack accounts & $37$ accounts\\
\hline
\end{tabular}
    }
}
\end{table}

\subsection{Study Approach}

\noindent 
To answer RQ1, we are focused on three important behaviors, including money transfer, account creation and contract invocation, mainly based on graph analysis. To answer RQ2, we perform an analysis of the account relationships to pinpoint bot candidates, and then perform behavior-level similarity comparison to identify real bot accounts. 
To answer RQ3, we first provide a taxonomy of the attacks found in the EOSIO platform. Then we further seek to explore the characteristics of different kinds of attacks with regard to their behaviors at the transaction-level. Based on the summarized behaviors, we have implemented a warning system to flag suspicious attacks and then perform manually verification.

\section{General Overview of EOSIO}
\label{sec:general}

We first conduct a comprehensive investigation of EOSIO, exploring \emph{money transfers}, \emph{account creation} and \emph{contract invocation}.

\subsection{Money Transfers}
\label{subsec:mfg}

\begin{figure}[t]
    \centering
        \label{subfig:MFG_times_dis}
        \includegraphics[width=0.48\textwidth]{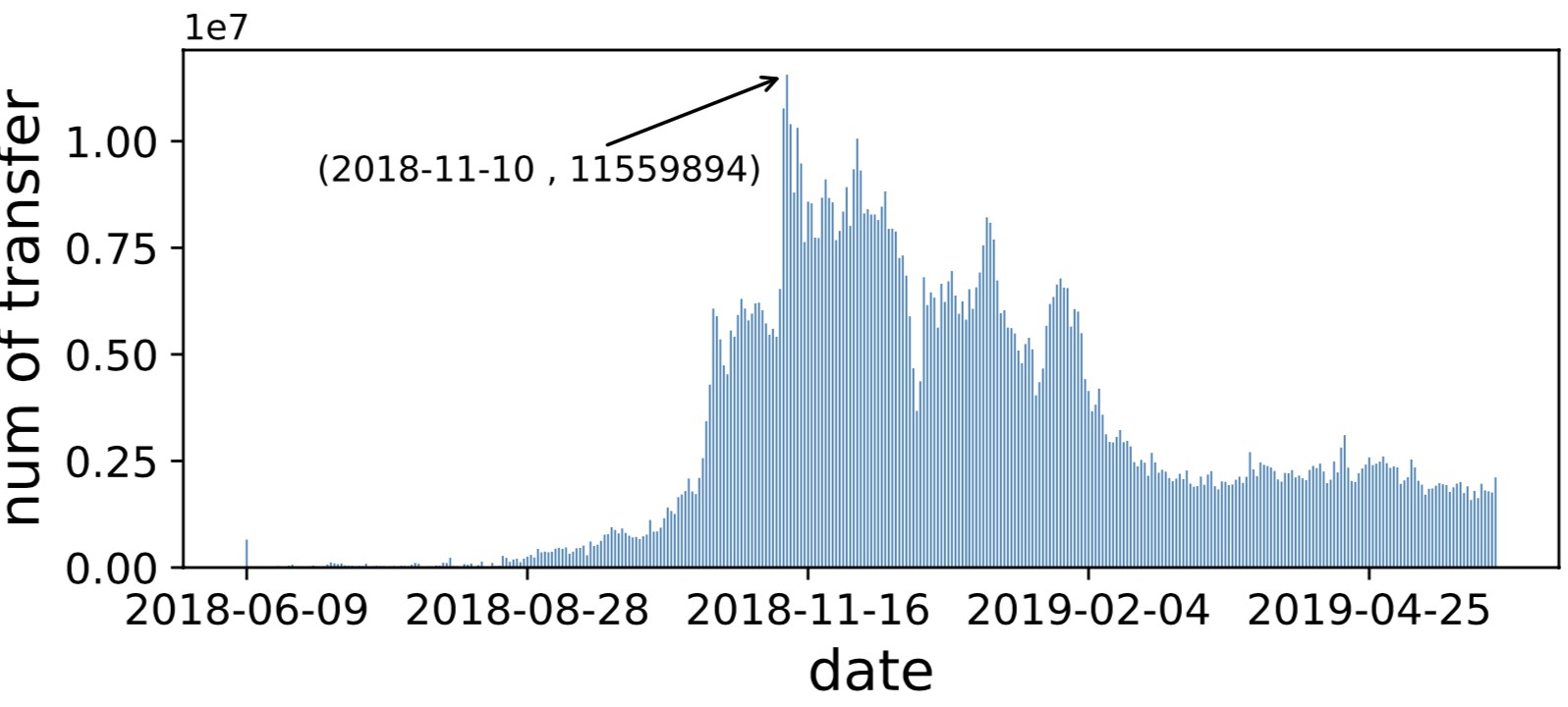}
        \caption{The Distribution of Money Transfer Over Time.}
\end{figure}

\subsubsection{Overview of Money Transfers}
\label{subsubsec:mfg_distribution}

\begin{figure}[h]
  \centering
  \includegraphics[width=0.48\textwidth]{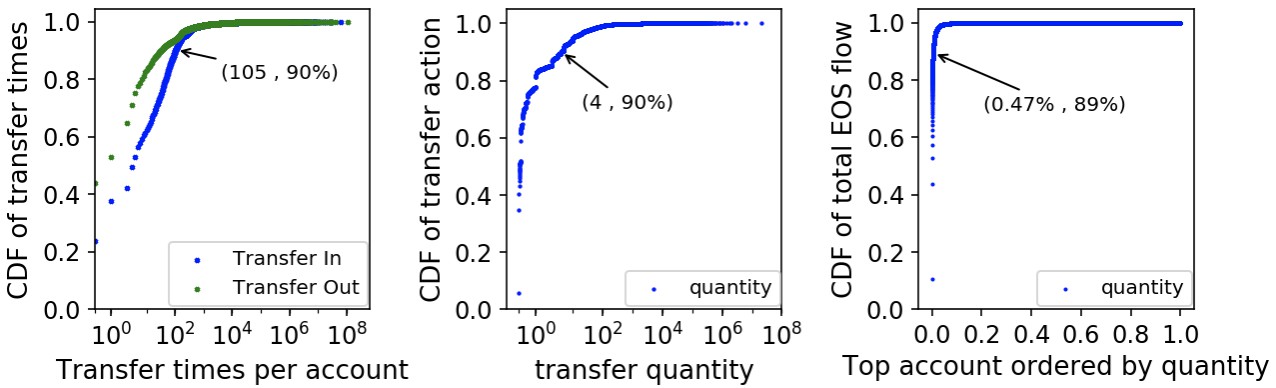}
  \caption{Distributions of Transfer Frequency and Quantity}
  \label{fig:transfer_distribution}
\end{figure}

The total number of money transfer transactions is over 1 billion ($1,055,690,229$), and the total number of EOS tokens being transferred is $15,190,552,483.9523$ EOS tokens. This represents  
over \$47 billion market value\footnote{We use the price up to {October 2019} to calculate the market value}
Fig.~\ref{fig:transfer_distribution} shows the number of transfers and accounts involved over time. We see that the number of transfers achieves its peak between 2018.11.09 to 2018.11.14. This is mainly due to the rising popularity of four gambling games (EOS Max, Dice, FAST and EOSJacks), as they cover 59.66 \% of the total transfers during that time. 

The distributions of transfer frequency and quantities are further shown in Fig.~\ref{fig:transfer_distribution}.
The number of accounts involved in transfers is $944,907$; in other words, $294,123$ ($23.74\%$ of all) accounts do not perform \emph{any} transactions.
Fig.~\ref{fig:transfer_distribution}(a) shows that the transfer frequency of the majority of nodes is below $100$, which implies that most accounts are not particularly active. One interesting detail is that $249,644$ accounts ($20.15\%$ of total) exclusively \emph{receive} EOS tokens but never transfer them out. 
Fig.~\ref{fig:transfer_distribution}(b) also shows the quantity exchanged within each transaction. For the 1 billion money transfers, most are small: over 90\% are under 4 EOS. We further analyze the total amount of money transfers for each account (cf. Fig.~\ref{fig:transfer_distribution}(c)), and find that $0.47\%$ of the accounts make up approximately $90\%$ EOS tokens being transferred, which is a typical \emph{Pareto effect}.

\begin{figure}[t]
\small
  \centering
  \includegraphics[width=0.48\textwidth]{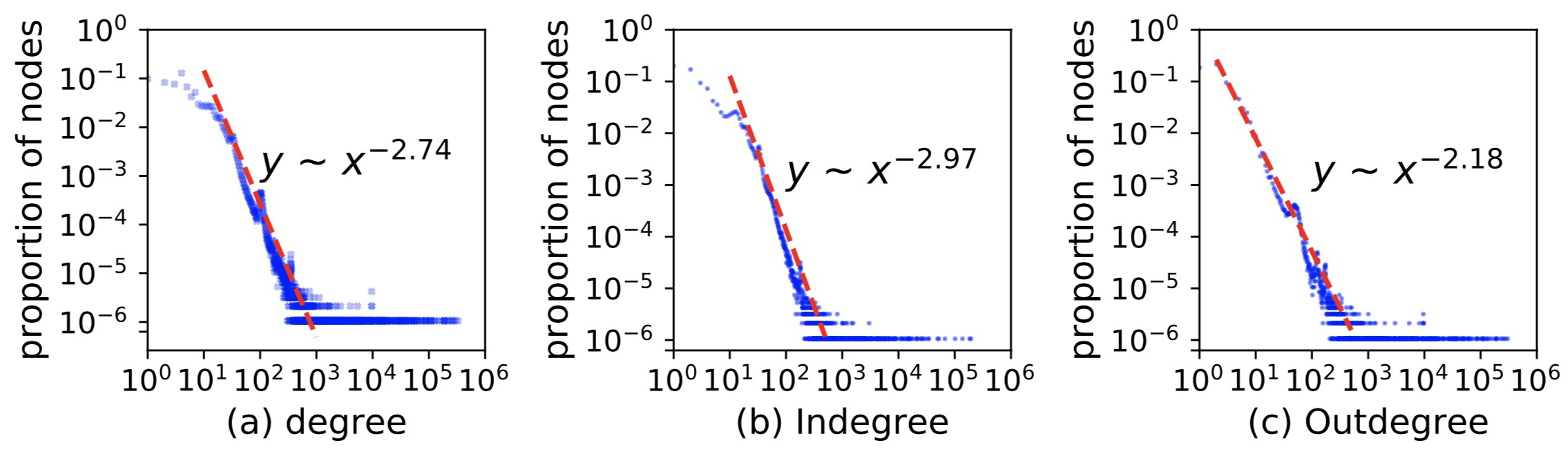}
  \caption{The Degree Distribution of EMFG.}
  \label{fig:MFG_distribution}
\end{figure}

\subsubsection{Graph Modeling of Money Transfers}
We next inspect which accounts perform transfers. To achieve this, we construct a graph from the set of $<from, to, value>$ transaction tuples.
However, this would ignore time information, which is important and necessary to capture abnormal behaviors. Therefore, we construct an Enhanced Money Flow Graph (EMFG) with timestamps, as follows:

\begin{equation*}
    EMFG=(V, E, D, w), E={(v_{i}, v_{j}), v_{i}, v_{j} \in v}
\end{equation*}

The order of the nodes in an edge indicates the direction of transferred money.
Each edge has at least one time attribute $d \in D$, $2018.06.09 \leq d \leq 2019.05.31$, indicating when the transfer occurs (UTC time). If account $v_{i}$ transfers multiple EOS to $v_{j}$ across multiple days, there will be more than one timestamp for edge $(v_{i}, v_{j})$.
$w: (E,D) \xrightarrow{}  \mathbb{R}^+$ associates each edge $(v_{i}, v_{j})$ on a particular day $d$ with the transfer quantity.

\begin{figure}[htbp!]
    \centering
    \subfloat[EMFG]{
        \label{subfig:CIG_num}
        \includegraphics[width=0.15\textwidth]{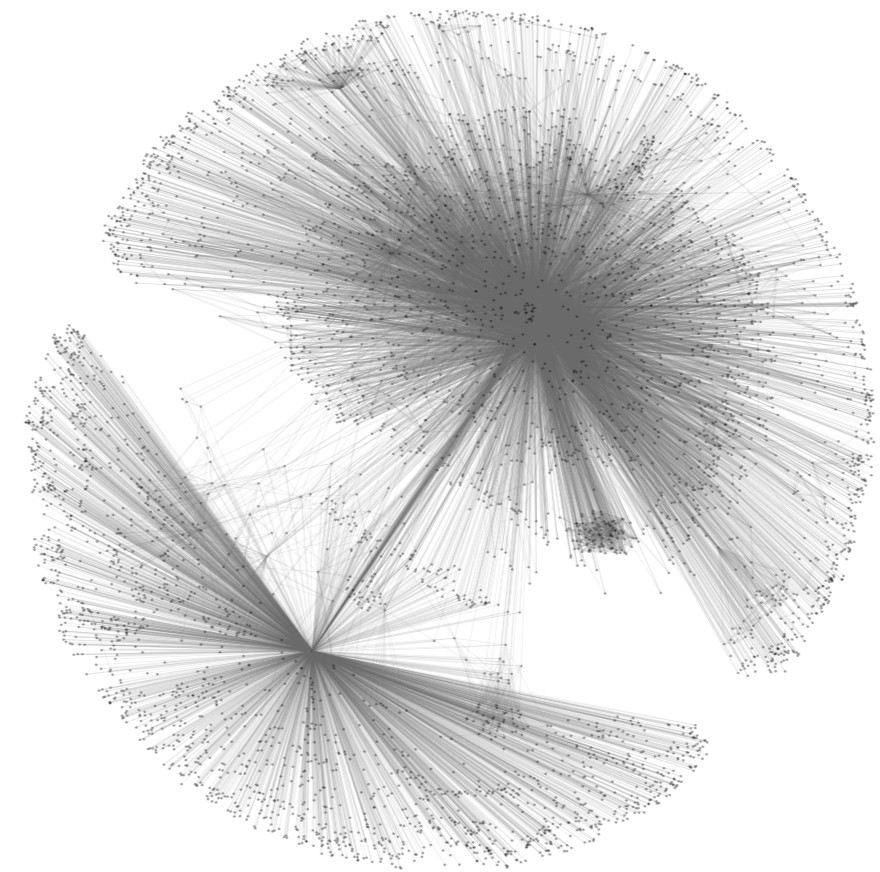}
     }
     \subfloat[EACG]{
        \label{subfig:CIG_target}
        \includegraphics[width=0.15\textwidth]{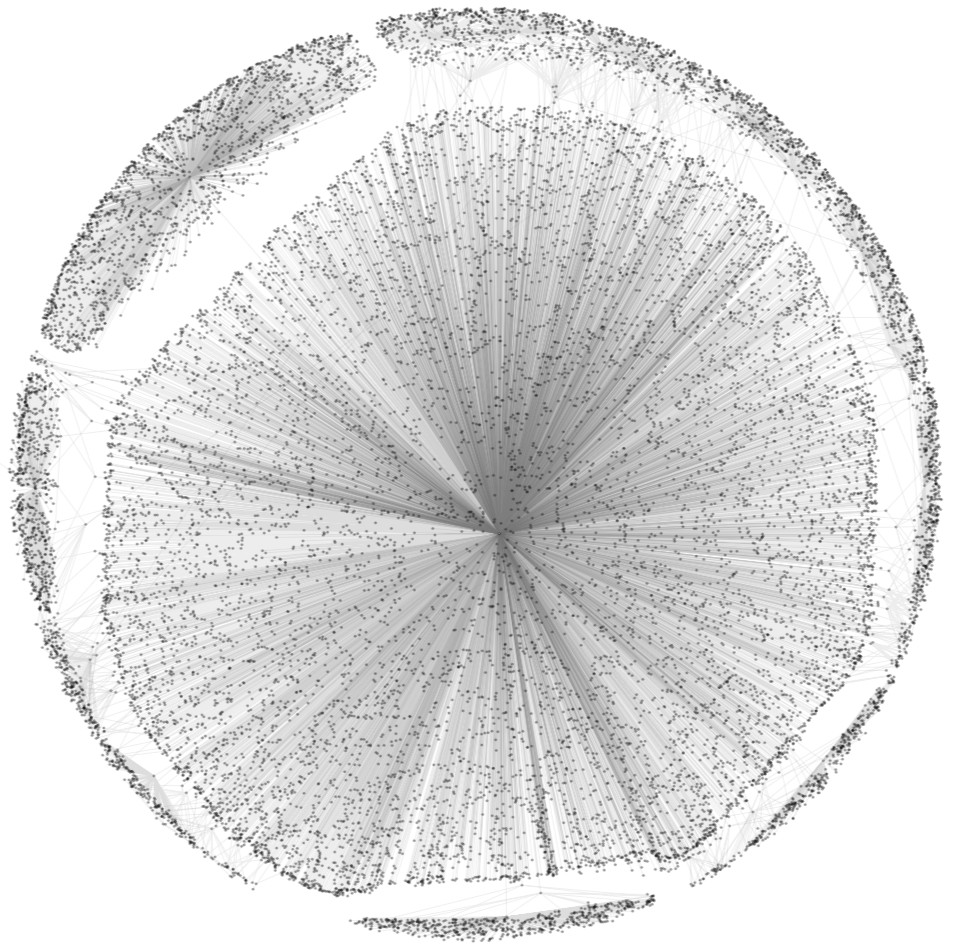}    
     }
     \subfloat[ECIG]{
        \label{subfig:CIG_target}
        \includegraphics[width=0.15\textwidth]{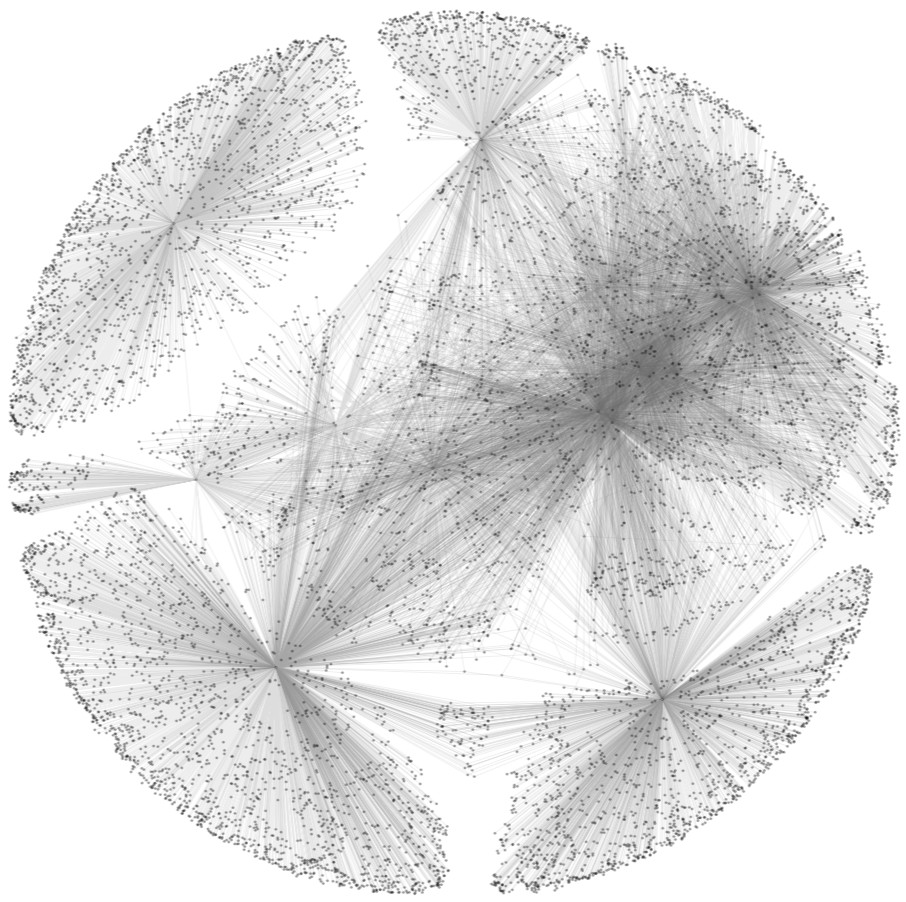}
     }
     \caption{Visualization of the constructed graphs.}
     \label{fig:visulization}
\end{figure}

\begin{table}[t!]
\centering
\caption{A Comparison of the graph metrics. Note that the numbers in \textbf{bold} are the metrics of EOSIO graphs, while the numbers in (brackets) are for Ethereum graphs~\cite{TingINFOCOM18}. }
\label{table:graph_metrics}
\resizebox{\linewidth}{!}{

\begin{tabular}{r|l|l|l}
\hline
\textbf{Metrics}   & \textbf{EMFG} & \textbf{EACG} & \textbf{ECIG} \\
\hline \hline
Clustering      &  $ \textbf{0.5049} $ ($0.17$) & $\textbf{0}$($0$) & $\textbf{0.2375}$ ($0.004$) \\
Assortativity   &  $\textbf{-0.3392}$ ($-0.12$)& $\textbf{-0.1539}$ ($-0.35$) & $\textbf{-0.1834}$ ($-0.2$) \\
Pearson         &  $\textbf{0.2448}$ ($0.44$)& \textbf{/}(/) & $\textbf{-0.00068}$($0.11$) \\ 
\hline
\# SCC          &  $\textbf{251,099}$($466,095$)& $\textbf{1,239,028}$($622,158$) & $\textbf{610,281}$($682,984$) \\
Largest SCC     &  $\textbf{693,632}$ ($1,822,192$)& \textbf{1}($1$) & $\textbf{688}$($84$) \\
\# WCC          &  $\textbf{1}$ ($81$) & $\textbf{3}$($22,260$) & $\textbf{3}$($4,088$) \\
Largest WCC     &  $\textbf{944,907}$ ($2,291,707$) & $\textbf{1,239,028}$ ($126,246$) & $\textbf{611,083}$ ($668,891$) \\
\hline
\end{tabular}
}
\end{table}

To measure the properties of the financial transactions, we apply some well-defined network metrics. 
The clustering coefficient~\cite{clustering-coefficient} measures the tendency that two nodes in a network cluster together; this is calculated using the approach of~\cite{fagiolo2007clustering}.
The assortativity coefficient~\cite{noldus2015assortativity} measures the preference for nodes to attach to others, i.e., a network is assortative when high degree nodes are, on average, connected to other nodes with high degree, and vice versa.
Assortativity is calculated based on~\cite{newman2003mixing}.
Pearson coefficient~\cite{Pearson} measures the linear correlation between two nodes.
We also extract the Weakly Connected Components (WCC), and Strongly Connected Components (SCC). 
Note that, \emph{we will use the same metrics to measure the graphs we construct for money transfer, account creation and contract invocation}.

\subsubsection{Results of EMFG Graph Modeling.}
Overall, there are $10,438,158$ edges and $944,907$ nodes. Fig.~\ref{fig:visulization} is a partial visualization of EMFG with $5,000$ nodes ($19,099$ edges) randomly selected.
The overall-degree, in-degree and out-degree distributions of EMFG are shown in Fig.~\ref{fig:MFG_distribution}, and all of them satisfy the power law distribution. 

The metrics for the constructed EMFG are shown in Table~\ref{table:graph_metrics}. For context, we compare against equivalent metrics for Ethereum, taken from~\cite{TingINFOCOM18} (note that although this study was performed in 2018, it still offers a useful comparison). 
First, we see the EMFG is a single WCC because of the specific account named \textit{eosio} (the first privileged account created by the system), which transferred to $163,937$ other accounts during the mainnet launch. This is different from Ethereum, which contains $81$ WCCs. The largest SCC contains $73.41\%$ of all nodes, which is similar to that of Ethereum ($86.59\%$). Like Ethereum, the fact that one WCC with multiple SCCs means that EOS tokens can flow from one SCC to another one but will not be transferred back (i.e., unidirectional transfer). 

The clustering coefficient of EMFG is $0.5049$, i.e., the likelihood of triadic closure between accounts, is almost three times as large as that of Ethereum ($0.17$~\cite{TingINFOCOM18}). 
The negative assortativity coefficient also implies that high-degree nodes are more likely to connect to nodes with lower degree. Such a phenomenal may come from the existence of accounts having one-to-many mapping relationship with others, such as exchanges, DApp games and some specific accounts like \textit{eosio} mentioned earlier.

We also identify the most \emph{central}accounts (measured by PageRank) in the EMFG. For the top-10 accounts, 3 accounts are exchanges and 4 accounts are gambling DApps. Comparing with the Ethereum study~\cite{TingINFOCOM18}, i.e., 8 exchanges and 0 gambling DApp, the composition of top DApps suggests that EOSIO is very much about speculation of value because nearly half of the top DApps are gambling games.
We further analyze the categories of DApp accounts, and find that the total volume of Gambiling DApps has occupied 78.88\% of the overall DApp volume, which further supports our observation.

\begin{framed}

\noindent \textbf{Findings \#1:} \textit{Most money transfers via EOSIO are small transfers (< 4 EOS). Although the overall ecosystem shows a promising volume, EOSIO is dominated by a small percentage of accounts (i.e., the top $0.47\%$ of accounts cover $90\%$ of the total volume).
Exchanges and gambling DApps dominate the transactions.}

\end{framed}

\subsection{Account Creation}
\label{subsec:acg}

\subsubsection{Overview of Account Creation}
\begin{figure}[t]
  \centering
  \includegraphics[width=0.45\textwidth]{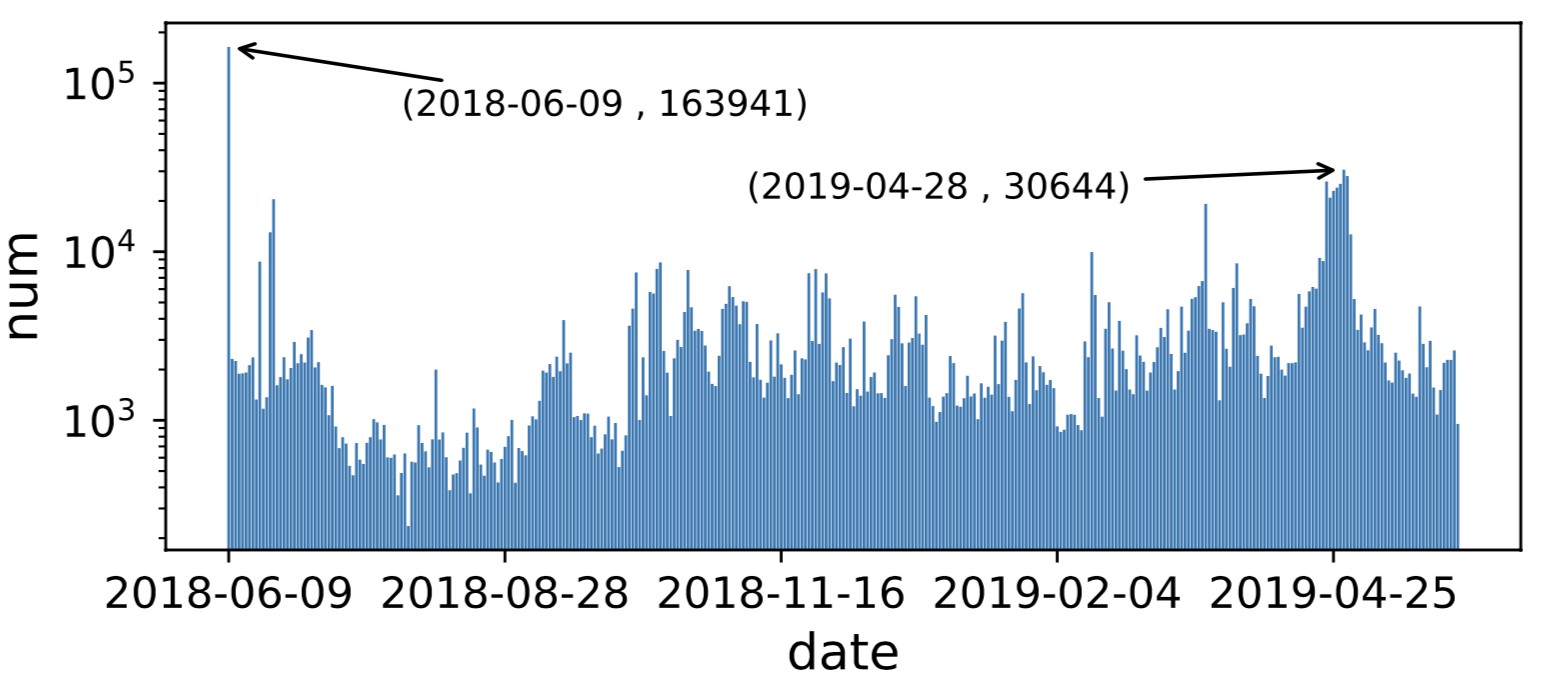}
  \caption{Distribution of Account Creation}
  \label{fig:ACG_distribution}
\end{figure}

We next inspect the account creation properties that we observe.
The times series of daily account creation is shown in Fig.~\ref{fig:ACG_distribution}.
There are several interesting peaks. The first stems from the token registration during the mainnet launch. Another peak, arising between $2019.04.23$ and $2019.04.29$, may seem a little strange though. $178,018$ accounts were created during this period, belonging to two wallet accounts: \textit{trxcashstart} and \textit{eostokenhome}. However, only a few accounts were directly created by them: $13,292$ for the former and $13,014$ for the latter.
The majority of the remaining accounts (from the $178,018$) were then indirectly created by the children of \textit{eostokenhome}. This ``spawning'' process is a bi-product of how accounts are created, hence we give further analysis in Section~\ref{sec:groupcontrol}.

This sporadic spawning of new accounts may suggest that not all are live and active. Hence, we next inspect if they, indeed, perform any transactions. 
Although creating an account expends system resources, \emph{the liveness of accounts on EOSIO is far below expectation.} 
Overall, $31.75\%$ accounts (i.e., $393,430$) are ``\textbf{silent}'' --- they never actively initiate any transactions with other accounts and never invoke any smart contracts. 
The proportion of silent accounts can therefore be used to reflect the liveness of EOSIO accounts.
Even those accounts that \emph{do} perform transactions, are rarely used. 
$66.24\%$ accounts performed 10 or fewer transactions. Such accounts inevitably undermine the prosperity of EOSIO.

\subsubsection{Graph Modeling of Account Creation}
Intuitively, the relationship formed during account creation activities can be modelled as a graph.
This is because new accounts can only be created by existing ones. Hence, we define and construct an Enhanced Account Creation Graph (EACG), as follows:
\begin{equation*}
    EACG=(V,E,D), E={(v_{i}, v_{j}), v_{i}, v_{j} \in v}.
\end{equation*}
An edge $(v_{i}, v_{j})$ indicates that account $v_{i}$ creates an account $v_{j}$. Each edge has a time attribute $d \in D$, $2018.6.9 \leq d \leq 2019.5.31$ (UTC time zone), indicating when the account is created.

\begin{figure}[h!]
    \centering
    \subfloat[CDF of invocation]{
        \label{subfig:silent_CDF}
        \includegraphics[width=0.18\textwidth]{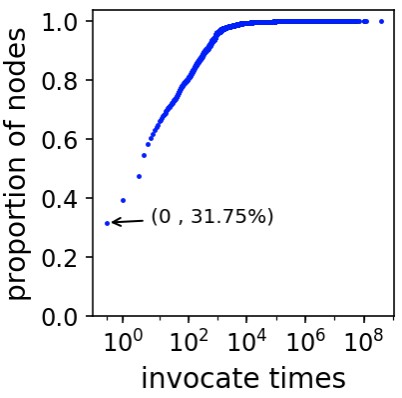}
     }
     \subfloat[Distribution of EACG]{
        \label{subfig:ACG_CDF}
        \includegraphics[width=0.19\textwidth]{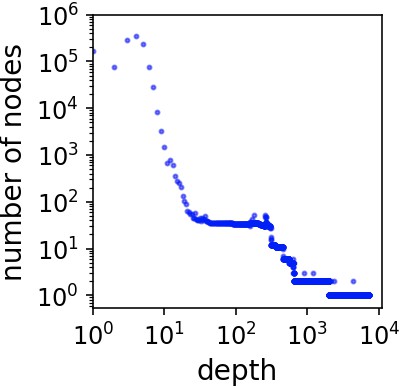}
     }
     \caption{Account Invocation and Creation.}
     \label{fig:ACGDistribution}
\end{figure}

\subsubsection{Result of EACG Graph Modeling.}
We compute our earlier graph metrics for the EACG, and present them in Table~\ref{table:graph_metrics}. The EACG consists of $1,239,027$ edges and $1,239,030$ accounts in total. 
In order to provide a more intuitive understanding of the graph, Fig.~\ref{fig:visulization}(b) gives a visualization for part of EACG. The account named \textit{imtokenstart} is selected as the root node, and the graph consists of $16,382$ nodes and their corresponding $16,381$ edges. This account is an example of a public service for account creation.

\begin{figure}[h!]
    \centering
    \subfloat[subtree of dogaigaohvwj]{
        \label{subfig:subtree1}
        \includegraphics[width=0.22\textwidth]{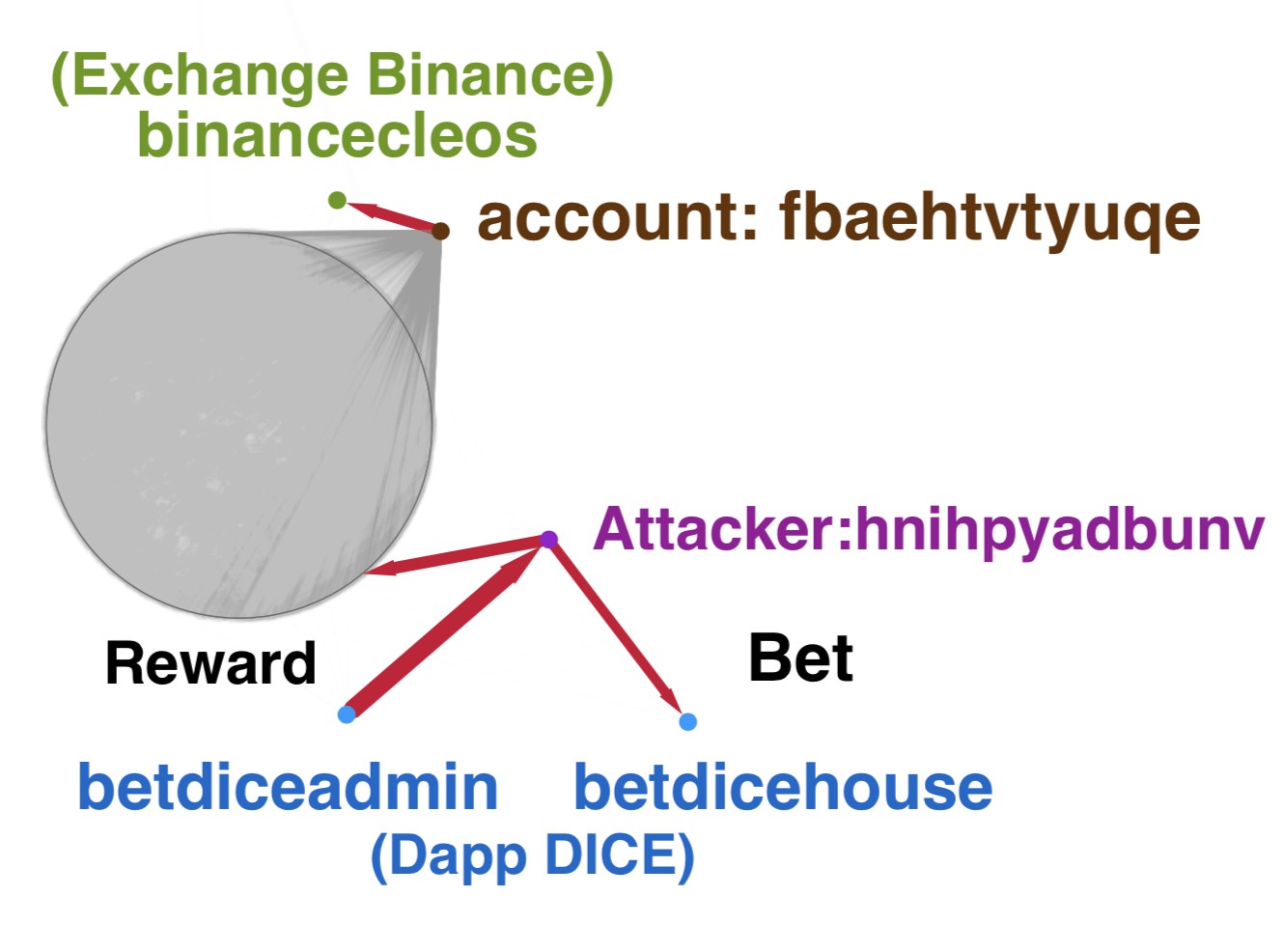}
     }
     \subfloat[subtree of chengcheng21]{
        \label{subfig:subtree2}
        \includegraphics[width=0.22\textwidth]{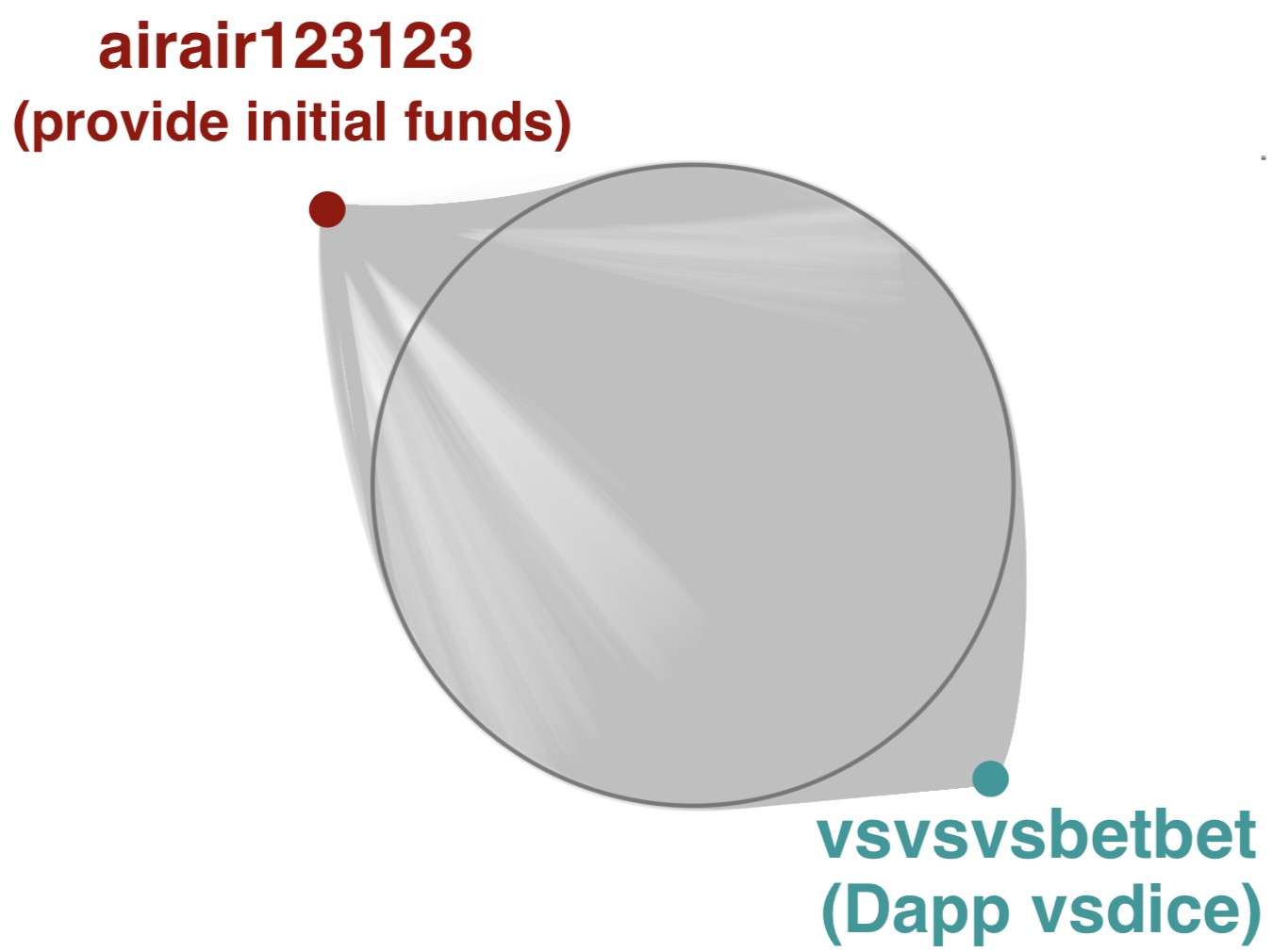}    
     }
     \caption{Visualization of the two anomalous paths.}
     \label{fig:subtree}
\end{figure}

We see that the EACG is a single tree when excluding a few isolated official accounts (\textit{eosio.prods}, \textit{eosio.null}), and the root node of the whole EACG, \textit{eosio}. 
Figure~\ref{subfig:ACG_CDF} shows the distribution of account trees by depth.
We see that EACG is a wide tree with a few deep paths. We see that from depth $651$ to depth $7,198$, there only exist \emph{two} paths (see Figure~\ref{fig:subtree}).
The first path has a root node named \textit{dogaigaohvwj}; it has a height of $7,195$, and was created to transfer illegal profits of a publicly known hacker account \textit{hnihpyadbunv}~\cite{fibos}. This attack was initially reported to include just $2,190$ sub-accounts~\cite{hnihpyadbunv}, yet we find this misses $5,005$ accounts. 
The second path has a root node named \textit{chengcheng21}; it has a height of $2,000$. Interestingly, a DApp named \textit{VSbet} received $95.30\%$ of the total EOS it transferred out. 
The activities of these accounts are quite suspicious. Almost all of them had transactions with \textit{VSbet}, and the total amount of EOS they transfer to \textit{VSbet} was almost identical to what they received; we later revisit this trend (see Section~\ref{sec:groupcontrol}).

Briefly, Table~\ref{table:ACG_importance} lists the top-5 central nodes.
All listed accounts are related to public services for account creation, allowing users to obtain new accounts. The existence of such services breaks the assumption that any new accounts are controlled by their creators. As we later revisit, this becomes an obstacle to identify attackers and bots on EOSIO.

\begin{table}[t]
\centering
\caption{Top-5 Nodes of EACG using Degree Centrality}
\label{table:ACG_importance}
\begin{tabular}{r|l|l}
\hline
\textbf{Account}& \textbf{Outdegree} & \textbf{Identity} \\
\hline \hline
eosio           & $163,946$ & Official account \\
senseaccount    & $50,531$  & Sense Chat \\ 
eosaccountwm    & $45,618$  & MEET.ONE \\ 
eostokenhome    & $35,908$  & ET Wallet \\
lynxlynxlynx    & $33,018$  & EOS LYNX \\
\hline
\end{tabular}
\end{table}

\begin{framed}

\noindent \textbf{Findings \#2:} \textit{Over $30\%$ of accounts are ``silient'' and have never initiated any transactions. The constructed EACG graph is a wide tree with several deep paths, while the outliers mainly belong to attack and fraudulent accounts. 
}

\end{framed}

\subsection{Contract Invocation}
\label{subsec:cig}

\begin{figure}[h]
    \centering
        \label{subfig:CIG_num}
        \includegraphics[width=0.45\textwidth]{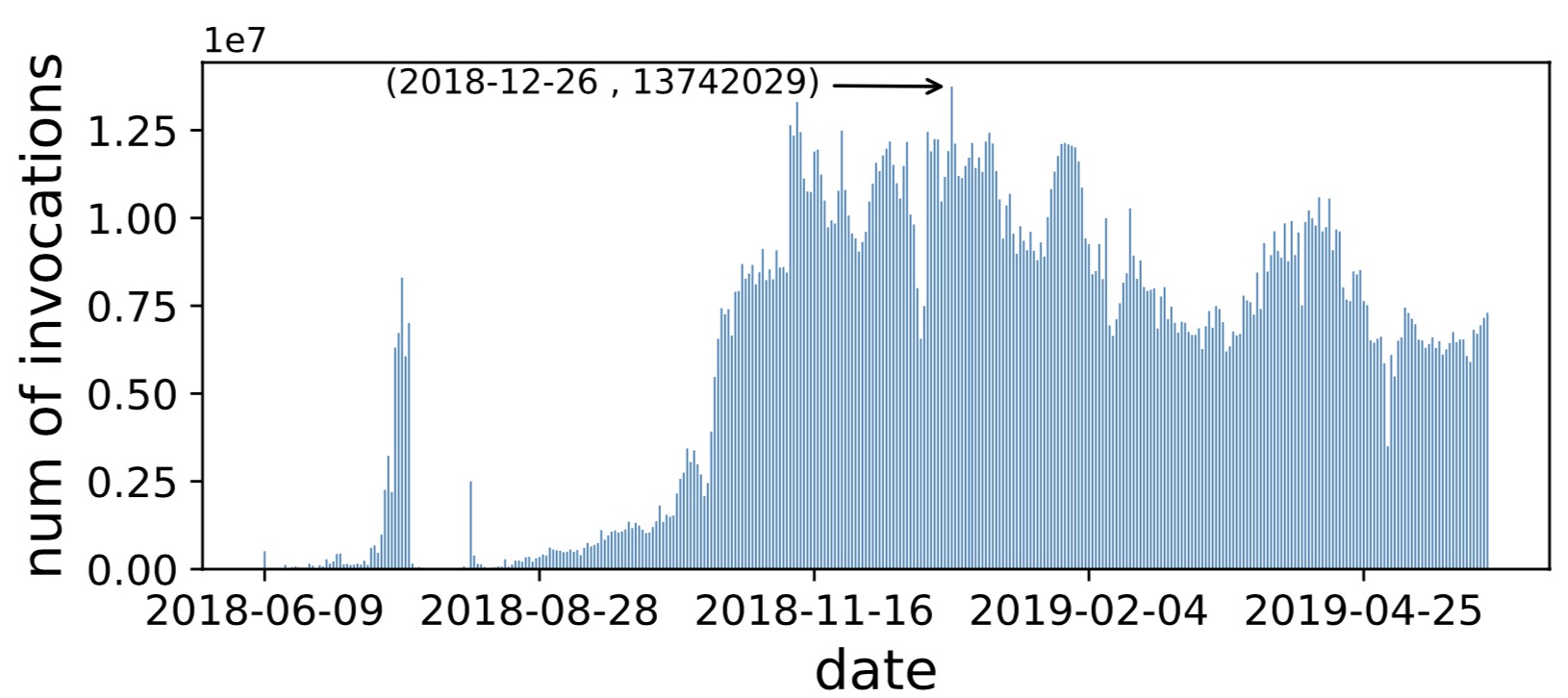}
        \caption{The Number of Invocation Over Time}
     \label{fig:CIG_overall_distribution}
\end{figure}

Finally, we inspect the contract invocation activities observed within EOSIO. 
There are 4,453 smart contracts in EOSIO, which have been invoked 2,140,945,703 times.
The time series of contract invocation is shown in Fig.~\ref{fig:CIG_overall_distribution}, which shows that creation has been stable after rapid growth in 2018.

\subsubsection{Graph Modeling of Contract Invocation}
Again, we can model contract invocations as a graph. Specifically, we define the Enhanced Contract Invocation Graph (ECIG), as follows:
\begin{equation*}
    ECIG = (V,E,D,A,f), E={(v_{i}, v_{j}), v_{i}, v_{j} \in v}.
\end{equation*}
An edge $(v_{i}$ and $v_{j})$ represents that an account $v_{i}$ invokes a smart contract $v_{j}$. 
To facilitate the identification of abnormal behavior, we annotate each edge with extra attributes.
Each edge has a timestamp attribute $d \in D$, $2018.6.9 \leq d \leq 2019.5.31$, indicating when the invocation occurs (in UTC time zone).
The name of the function for each invocation is also recorded as $a$, $a \in A$. $f: (E,D,A) \xrightarrow{} \mathbb{Z}^+$ assigns each edge $(v_{i}, v_{j})$ with a particular day $d$ and given action $a$ the number of invocation.

\subsubsection{Results of ECIG Graph Modeling} 
Overall, there are $4,313,739$ edges and $611,085$ nodes. Fig.~\ref{fig:visulization}(c) is a partial visualization of the ECIG with $10,626$ nodes (and $19,079$ edges) being randomly selected.
We compute our earlier metrics, and present the results in Table~\ref{table:graph_metrics}. Because of the existence of the system contract, the largest SCC of EOSIO is larger than that of Ethereum, although there are only $4,453$ contracts being invoked on EOSIO.

Table~\ref{table:CIG_importance} lists the top-5 most important nodes in the ECIG.
An intriguing observation is that no exchanges are present in Table~\ref{table:CIG_importance}. This is quite different from Ethereum (7 reported by a recent study~\cite{TingINFOCOM18}). Notice that the number of invocations by these nodes are quite high compared to their degrees. Combined with the observation from Figure~\ref{subfig:silent_CDF}, we can confirm that most accounts rarely invoke smart contracts. Obviously, a small set of accounts are much more active than the others: the top 0.01\% of accounts perform 80 \% of contract invocations.
Our manual investigation suggests that there are many bots on EOSIO, as we will explore in Section~\ref{sec:groupcontrol}.

\begin{table}[htbp!]
\centering
\caption{Top-5 Nodes of ECIG using Degree Centrality}
\label{table:CIG_importance}
\resizebox{\linewidth}{!}{
\begin{tabular}{r|l|l|l}
\hline
\textbf{Account}& \textbf{Degree}   & \textbf{\# of Invocation}   & \textbf{Identity} \\
\hline \hline
eosio           & $383,590$         & $8,925,999$   & Official account \\
betdicelucky    & $109,151$         & $48,694,997$  & DApp Dice \\ 
pornhashbaby    & $107,417$         & $64,072,844$  & DApp Hash Baby\\ 
hashbabycoin    & $96,826$          & $31,380,78$   & DApp Hash Baby \\
endlessdicex    & $93,469$          & $23,955,881$  & DApp Endless Game \\
\hline  
\end{tabular}
}
\end{table}

\begin{framed}

\noindent \textbf{Findings \#3:} \textit{Although there are only $4,453$ contracts in the platform, EOSIO has a large number of invocations. 
A small number of accounts are much more active than the others. The top 0.01\% of accounts perform 80\% of contract invocations, while the majority of accounts rarely invoke contracts. 
}
\end{framed}

\section{Characterizing Blockchain Bots}
\label{sec:groupcontrol}

It is well-known that online activities are impacted by bots~\cite{InternetBot}.
We posit that these might heavily also impact the EOSIO ecosystem, and therefore we next investigate \emph{how many activities are driven by bots, and what are their incentives (purposes)}.

\subsection{A Primer on Bots}

We define \textbf{bot-like account} as those operated by machines (e.g., programs).
These are already known to exist in EOSIO, e.g., the famous DApp team ``pornhashbaby'' has reported 8 groups of bots, and each group has hundreds to thousands of accounts~\cite{eos-hall-of-shame}.
These past findings have shown that bot accounts usually operate in groups, and have repetitive behavioral patterns.
In this paper, we refer to a group containing \textbf{bot-like accounts} operated by a controller as a \emph{bot-like community}. 
Note, the bot accounts may be created by the same controller account (in the same sub-tree of EACG) or by a number of different accounts (across different sub-trees of EACG). We will have two complementary approaches to detect them.

\subsection{Preliminary Observations}
\label{sec:bots:preliminary}

To help distinguish bot accounts from other normal accounts (especially public account creation services), we perform a preliminary study of bot activities. 
We do this by harvesting a number of ground-truth bot accounts from~\cite{eos-hall-of-shame}, as well as a well-known blockchain security company. 
From this, we gather $63,863$ EOSIO bot accounts flagged by existing efforts. 
The accounts were created within 21 bot-like communities. 
To further differentiate these accounts from normal ones, we label $229,907$ normal accounts that belong to 25 public contract services (e.g., official account, exchanges, and pocket) as a white-list for comparison.
We then manually inspect these two groups and make several observations from two perspectives: \textit{community-level} and \textit{account-level}.

\subsubsection{Community-level Observation}
First, controller accounts usually create a large number of children accounts (bots). For the flagged 21 controller accounts, the out-degree of them in the EACG varies from 108 to 15,025. In contrast, the average out-degree of the EACG graph is 30, while the median is only 1. Thus, we argue that a ``shortlist'' of potential bot controllers could be identified using the EACG graph. Note that this is \emph{not} definitive though --- many non-bots also have high out-degrees. 
Hence, we turn to our second observation, where we find that accounts belonging to the same bot-like community tend to share similar behaviors. For instance, they might perform transactions on the same day using the same contract. We posit that this similarity could be used to automatically group bots belonging to the same controller.

\begin{figure}[t]
\small
  \centering
  \includegraphics[width=0.9\linewidth]{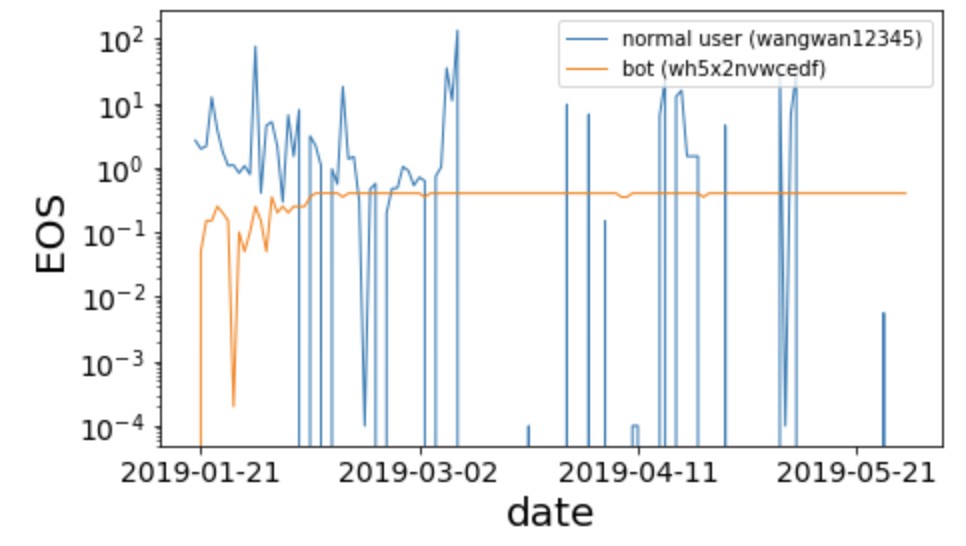}
  \caption{An Example of difference between bots and normal users in EOS transfer}
  \label{fig:feature_example}
\end{figure}

\subsubsection{Account-level Observation}
We further analyze the bots at a per-account level. As bot accounts are controlled by machines, their behaviors usually have regular patterns. Figure~\ref{fig:feature_example} shows an example of a labelled bot account and a normal account. The bot account has relatively small but very frequent incoming transfer activities from February 2019 to May 2019. By analyzing its activities, we find that it is performing Click Fraud. Each day it transfers a few EOS to eosvegasjack (an account that belongs to a gambling game) several times. After each transfer, the bot instantaneously gets its money back. The time interval between each transfer is also regular (usually three hours).
Thus, we believe that the bot accounts could be classified based on their behavior patterns, including the active time, frequency and volume of transfers, etc.

\subsection{Detecting bots from community-level} \label{ssec: community}

Based on the aforementioned patterns, we next devise a simple algorithm capable of identifying bot-like communities. 

\subsubsection{Algorithm Design}

Our approach combines both \emph{account relations} and \emph{behavioral similarities} to flag suspicious bot accounts.
The algorithm has two stages. In the \emph{first} step, we extract all accounts that have created in excess of 30 new accounts (which is the average out-degree for nodes in the EACG). 
This provides a shortlist of accounts that \emph{may} be spawning many accounts for use as bots. 
However, it is not necessarily accurate, as we have already identified legitimate services that also spawn many accounts.
Hence, the \emph{second} step computes the similarity between accounts on the shortlist to identify groups that operate in similar ways. 
To measure this similarity, we compare accounts across two dimensions. 

The first dimension is the \textbf{invocation time and frequency}, i.e., the time and frequency of invoking smart contracts and transferring money.
Specifically, for each account, we summarize its money transfer actions and smart contract invocations in a feature vector. For each account on day $i$, we calculate the frequency of its money transfer actions, and its smart contract invocations, respectively. As we have collected all the action records since the launch of the EOSIO mainnet (which lasts 357 days), the behavior of each account can be represented in a 714 dimension feature vector $\vec t_{i}$  (357 for the money transfer actions, and 357 for the contract invocations).

The second dimension is the \textbf{contract target} of the transactions. This is because accounts belonging to the same bot operator might invoke the same smart contracts. There are currently 4,453 unique smart contracts in the EOS ecosystem. Thus, for each account $i$, we represent its targets in a 4,453 dimension vector $\vec s_{i}$. Each dimension represents the number of times that the account has invoked the corresponding smart contract.

The above provides two dimensions for which we can compute similarity between potential bot accounts. 
We use \emph{cosine distance} to measure the similarity between feature vectors. Considering there are a number of accounts in a potential bot-like community, we further calculate a \emph{group similarity} for each of them as follows:
    (1) For a community with N non-silent nodes,\footnote{We have removed ``silent'' nodes, as such nodes have no actual behaviors to measure.} we calculate the feature vectors $\vec s_{i} \in S $ and $\vec t_{i} \in T $ for each node.
    (2) We further calculate the median vector $\vec m$, $\vec m=\frac{1}{N}\sum_{j=1}^N \vec v_{j}$  for vectors in $S$ and $T$.
    (3) We then measure the average distance between each vector and median vector. $\overline{dist} = \frac{1}{N}\sum_{j=1}^N cosineDist(\vec v_{j}, \vec m)$. As a result we get the overall $\overline{dist_{S}}$ and $\overline{dist_{T}}$ for S and T. Note that $\overline{dist} \in [0,1]$, where larger values means less similarity.

\begin{figure}[t]
\small
  \centering
  \includegraphics[width=0.48\linewidth]{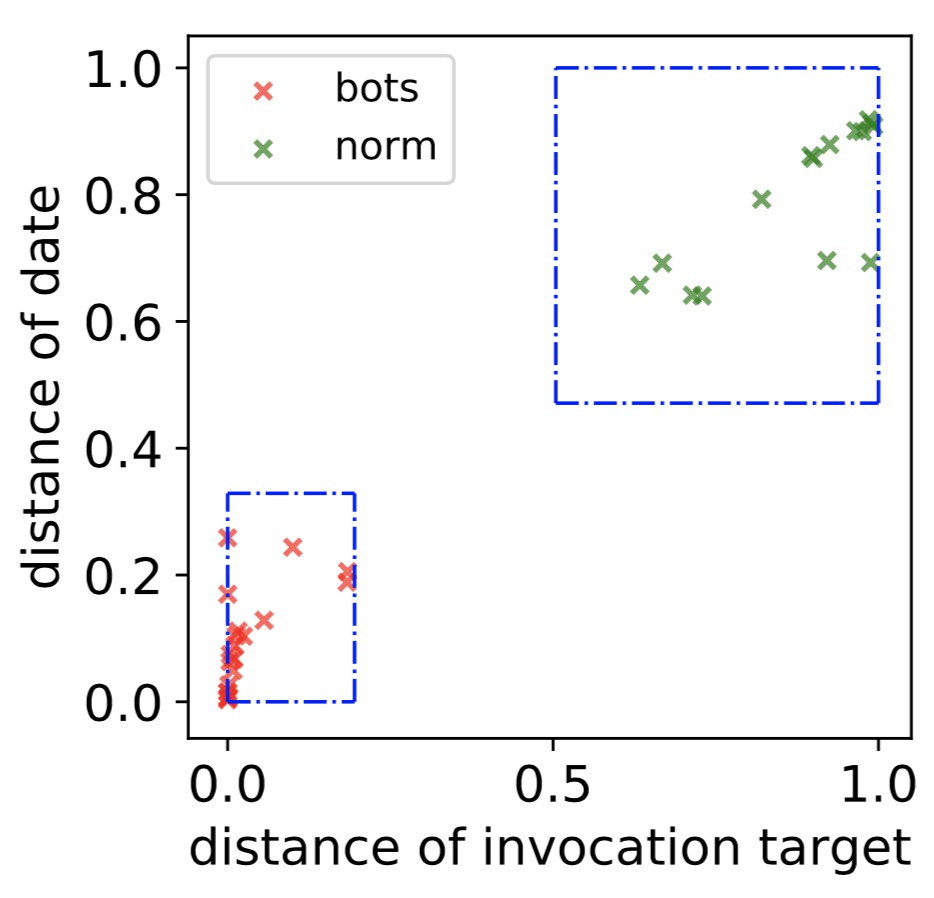}
  \caption{Bot-like Communities VS. Normal Communities.}
  \label{fig:bots_distance}
\end{figure}

\subsubsection{Defining the Similarity Threshold.}

\begin{table}[t]
\centering
\caption{Bot-like Communities VS. Normal Ones.}
\begin{tabular}{l|l|l|l}
\hline
                        & Metrics       & Bots & Normal \\ \hline
\multirow{2}{*}{Time Features}   & Average       & 0.09 & 0.78   \\ \cline{2-4} 
                        & Std.Deviation & 0.08 & 0.10   \\ \hline
\multirow{2}{*}{Target Features} & Average       & 0.03 & 0.87   \\ \cline{2-4} 
                        & Std.Deviation & 0.05 & 0.12   \\ \hline
\end{tabular}
\end{table}

The above allows us to compute the similarity between accounts. The next step is to define an appropriate threshold of similarity that warrants a community of accounts being classified as bots. Here we take an empirical approach, relying on our ground-truth dataset of bots and normal accounts (from Section~\ref{sec:bots:preliminary}).
Note that this covers 21 bot communities (63,863 accounts) and the 25 normal ones (314,277 accounts). 
Figure~\ref{fig:bots_distance} presents the similarity across these 46 groups. Groups that fall on the lower left have tightly correlated behavior. Specifically, the x-axis represents the average distance of the action target feature vectors, and y-axis represents the average distance of the action time and frequency vectors.

Note that by Chebyshev’s Theorem, even for non-normally distributed variables, at least $ 88.8\% $ of cases should fall within properly calculated 3$\sigma$ intervals. 
Hence, we use the similarity score ranges between -3$\sigma$ and +3$\sigma$ of the average as our threshold to filter bots (see the rectangular frame in Fig.~\ref{fig:bots_distance}). 
Any groups of accounts that fall into this range are deemed to belong to the same bot groups.

\subsubsection{Analyzing the Public Key.}
We also observe that some users re-use the same public-private key pair across multiple accounts. Hence, as a final step, we group all the flagged bot-like accounts sharing the same public key into a community.

\subsubsection{Results.}
Using the above techniques, we identify $351,560$ bot-like accounts, which cover $28.37\%$ of all accounts in the EOSIO platform, and belong to $2,363$ communities.
The number of accounts in the identified bot communities varies from $30$ to $15,025$ .

\subsection{Detecting bots from per-account level}
\label{sec:bot-isolated}

Bot accounts are not always created by the same controller account. 
For example, DApps like \textit{MEET.ONE} and \textit{Infinitowallet} sometimes provide free accounts to attract users. Thus, bonus hunters would take advantage of this opportunity to create free bot accounts. 
In this case, the bot accounts may reside within the ACG trees mixed up with normal accounts. To overcome this, we must also perform per-account detection.

\subsubsection{Algorithm Design}
To differentiate accounts within the same ACG tree, we formulate the problem as a binary classification task. 
Based on the graphs we constructed, we summarize 11 key account features in Table~\ref{table:feature}.
We describe some representative features here.
For EACG, the \textit{ACG depth} describes the depth of the account in EACG. Bots appear to be deeper in the tree. For EMFG, as illustrated in Fig.\ref{fig:feature_example}, bots usually receive EOS more frequently than normal accounts. So we use features like \textit{transferIn std} to describe the standard deviation of transfer volume over time. The amount of EOS per transfer for bots is also much lower than that of normal ones. Thus, we summarize the \textit{volume per transferIn} as a feature. For ECIG, it appears that bots are more active than normal accounts, i.e., the active day is 74.6\% (bots) VS. 13.4\% (normal) on average. In other words, bots are more active than normal accounts. We use \textit{activate time} to describe the activity. The average values of aforementioned features in the labeled dataset are shown in Table~\ref{table:feature}.

\begin{table}[t]
\centering
\caption{Features we used to classify bot accounts.} 
\label{table:feature}
\setlength{\tabcolsep}{7mm}{
\resizebox{\linewidth}{!}{
\begin{tabular}{l|l|l}
\hline
     feature & bots (avg) & normal (avg)\\
\hline
\hline
     ACG depth & $4.3$ & $1.8$ \\
\hline
     transferIn std & $40.8$ & $954$\\
\hline
     transferOut std & $57.3$ & $1267.3$ \\
\hline
     volume per transferIn & $2.2$ & $104$\\
\hline
     volume per transferOut & $6.2$ & $553.1$\\
\hline 
     transfer target num & $5.9$ & $12.6$ \\
\hline
     invoke contract num & $16.1$ & $3.2$ \\
\hline
     invocation num & $518.5$ & $2312.7$ \\
\hline
     invocation std & $6.2$ & $40.3$ \\
\hline
     activate time & 74.6\% & 13.4\% \\
\hline
    siblings in same day & $1724.1$ & $117024.4$ \\
\hline
\end{tabular}
    }
}
\end{table}

Using these features, we train a random forest to classify accounts into bots vs. non-bots.
After performing a random shuffle, we split our labeled dataset to training (80\%) and test (20\%) sets. After doing grid search for the best parameters, we finally get $99.56\%$ accuracy on the test set. We will release the benchmark and the classifier to the community.

\subsubsection{Results}
We use the classifier to identify bot accounts in the remaining $486,328$ accounts.
We discover $29,448$ new bots. 
When merged with the community-level results, \textbf{we identify  381,008 bot accounts in total}. These accounts have invoked $192,330,821$ times, and the amount of transferred EOS is $639,517,282$.

\subsection{Validating Bot Accounts}

Before continuing, it is important to validate that the above approaches effectively identifies bot accounts. We use three methods. 
\emph{First}, we compare our results against the ground-truth bot accounts (see Section~\ref{sec:bots:preliminary}), to find that we gain 100\% accuracy --- all $63,956$ ground-truth bot accounts are identified by our techniques. In fact, by only using the random forest approach on the per-account level, we could achieve an accuracy of 99.51\%.
\emph{Second}, we have reported a number of bot communities to the DApp developers and one anonymized blockchain security company, and all of them were confirmed as bots. 
\emph{Third}, we randomly select 100 bot-like account communities for manual examination. We are confident that over $88\%$ are indeed bots, and we are able to know their purposes (see Section~\ref{sec:botclassify}). For a further $12\%$ of the bot-like accounts, although we cannot reverse engineer what they are doing, we are able to identify some bot-control clues, e.g., the accounts have similar names, a large number of accounts were created at the same time, etc. 
We will release our detection result.

\subsection{Applications of Bot Accounts}
\label{sec:botclassify}

We next examine the purposes of these bot accounts. We manually sample 100 bot communities, and examine their actions, identifying 4 major categories, which we subsequently formulate automated methods to detect.
Table~\ref{table:bot-summary} provides a summary of these categories, highlighting their significant impact of the wider ecosystem. Below, we discuss each category.

\subsubsection{Bonus Hunters.} \label{sssec: bonus_hunter}
We find many bot groups performing \emph{bonus hunting}. This involves exploiting incentives in DApps, in order to gain profit. This is because EOSIO DApps often offer incentives to attract users, including:
(1) \emph{Login bonus}: a reward given to users for logging into DApps daily. For example, DApp BingoBet offers users one free lucky draw every day. Users can get free EOS from 0.0005 to 50.0.
(2) \emph{Free tokens}: some DApps hand out tokens to attract active users. For example, PRA Candybox offers each user a limited chance to get free tokens. 
In general, each account can only get a limited number of tokens, so by creating multiple bots, a controller can make more profit. 
(3) \emph{Invitation rewards}: some DApps give rewards for inviting friends. For example, Hash Baby offers 5\% of its total tokens to those who invite friends. Thus, bots could create a large number of accounts and invite them in order to gain profit. 

Although we have manually identified many bonus hunters, it is non-trivial to automatically identify them. 
Here, we propose a semi-automated but effective approach.
One feature of bonus hunter bots is that they invoke some specific contracts with a high frequency, or they will invite other ``similar'' accounts to invoke the contracts. 
Thus, we first rank the bot-like accounts by their frequency of contract invocations, and then check the most popular DApps they invoke. For the top DApps, we then check to confirm whether they offer the aforementioned incentives. As a heuristic, we assume that bot communities targeting these apps are likely bonus hunters. 
To gain an upper estimate of how many bot communities are bonus hunters, we compute the proportion of money hunting actions for each account. We consider anything above 50\% to be indicative of a likely bonus hunter. Through this, we identify 158,147 accounts belonging to this category. It accounts for 36.55\% of all bot-like accounts. These accounts contribute the second largest number of invocations (30.43\%).

\subsubsection{Click Fraud.} 
The second largest  use of bots is for \emph{click fraud}. These accounts create ``fake'' traffic in order to promote the ranking of certain DApps.
One major characteristic is that \emph{the amount of EOS they transfer to DApps is almost identical to what they receive from the same DApps}. 
We also observe that these accounts are quite ephemeral, i.e., they become active for a few days and then become ``silent''.
This results in the rank of the targeted DApp being boosted temporarily during the active days. 

\begin{figure}[t]
\small
  \centering
  \includegraphics[width=0.8\linewidth]{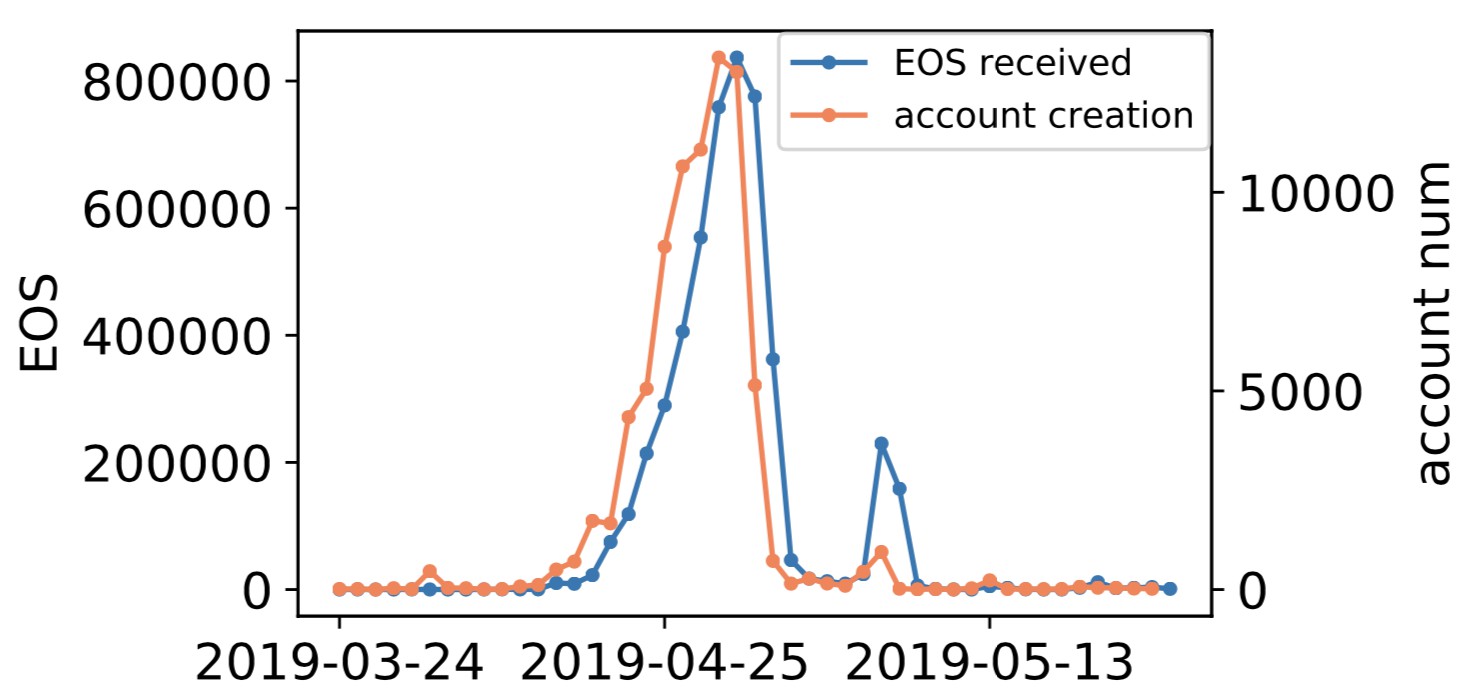}
  \caption{An Example of the Click Fraud (\texttt{EOS Global}).}
  \label{fig:bots_traffic}
\end{figure}

The most extreme case is the \emph{EOS Global}, which ranks 3rd among all DApps on 2019-4-27 (on DappTotal). Using our EMFG, we find that account ``egtradeadmin'' (belonging to \emph{EOS Global}), ranks \#1 among all accounts using PageRank. However, the transaction behaviors of ``egtradeadmin'' are typically bot-like. To highlight this, Fig.~\ref{fig:bots_traffic} shows the money transfer and account creation behaviors of \emph{EOS Global}. 
A significant peak can be witnessed between 2019-04-23 and 2019-04-30. The accounts contributing were flagged as bot-like accounts using our approach, as since their creation, they have transferred a great deal of EOS to the \emph{EOS Global} DApp, and received almost identical Figure EOS from \emph{EOS Global} in return.

We define a heuristic to classify bot-like accounts in this category. We simply check their balance (transfer in vs. transfer out) to identify whether they are suspicious. 
Accounts with a similar ratio (over $95\%$) are deemed to be performing click fraud.
Note that, we have filtered bot-like communities that have very few transfers (e.g., <10 EOS) to remove potential false positives.
We identify $139,271$ bot-like accounts belonging to this category. They have created $80,035,916$ invocation, with an amount of $236,607,485$ EOS.

\subsubsection{DApp Team Controlled Accounts}

We also observe that some bot-like accounts are controlled by DApp teams. 
By analyzing the behaviors of such accounts, we see that they are mainly used for: 
(1) \emph{Debugging:} These bots are used to run automated product quality assurance tests within the DApp. For example, account ``zhaojingdong'' has created 2,001 bot-like accounts, and these accounts are used to invoke the debug interface of the smart contract ``fastwinhold1'' (fastwinhold1::debug), with 19,299 invocations. The ``fastwinhold1'' is a smart contract that belongs to \emph{Dapp Fast}. 
(2) \emph{Interacting with users:} These accounts are created to support user-based activities within DApps.
There are also other cases where DApp Games cannot find enough real players (especially during cold start), so they generate bots to interact with real users. 
(3)  \emph{Control:} These bot accounts are mainly used to either vote for themselves or run DApps of their own. These accounts have the same public key with DApp accounts, e.g. ``eosiomeetone'', ``eosasia11111''. As the EOS inherits DPoS and only 21 block producers can be elected, these accounts are used to perform self-voting.

As we have collected the DApp info, Team Controlled Accounts can be verified easily. We have identified 19,567 accounts (5.14\%) belonging to this category. 
Note that these accounts invoke smart contracts with high frequency (52,307,259 times in total), which represents 27.20\% of the overall bot-like invocations.

\subsubsection{Account Sellers.}  \label{sssec:account_sellers}
Account Sellers involve creating large numbers of accounts with names that are perceived to have potential future value, as names are unique in EOSIO and cannot be reused. 
Account Sellers therefore bid for account names from the EOS authority, and then sell them for a higher price. The public key of the accounts are identical and only after they are sold out will they change their keys. 
We identify these account sellers by examining the websites of each DApp, as they will advertise their ability to sell accounts. 
From this, we have identified 15,496 accounts in total.

\subsubsection{Others.}
12.74\% of the bot-like accounts (48,527 accounts, 248 communities) remain uncategorized, as we cannot identify generalizable traits that might reveal their motivation.  
That said, we are nevertheless confident that they are bots.
From this group, we manually sample 11,948 bot accounts (132 communities), and observe several clues that give weight to this confidence:
(1) A large portion of these bot communities (35.47\%) have confusingly similar names. For example, accounts created by gotolab12345 are named things like gotolabms221, gotolabms222, while for normal accounts, we rarely see this.
(2) The account creation time of most of the bot communities (64.53 \%) is extremely intense, making normal human involvement unlikely, \eg account aaa1bbb1ccc1 created 523 accounts in a few hours.

\begin{table}[]
\centering
\caption{A Summary of the bot-like accounts.}
\label{table:bot-summary}
\resizebox{\linewidth}{!}{

\begin{tabular}{l|l|l|l}
\hline
Category       & \# Accounts (\%)  & \# Invocations (\%)  & \# Volume (EOS) \\ 
\hline \hline
Bonus Hunter   & 158,147 (41.51\%) & 58,521,684 (30.43\%) & 57,460,802(8.99 \%)                \\ 
Click Fraud    & 139,271 (36.55\%) & 80,035,916 (41.61\%)   &  236,607,485(37.00 \%)              \\ 
Dapp Team      & 19,567 (5.14\%)   & 52,307,259(27.20\%) & 322,731,329(50.46 \%)               \\ 
Account Seller & 15,496 (4.67\%)    & 327,192 (0.17\%)     & 3,531,674(0.55\%)                \\ 
Others         & 48,527 (12.74\%)   & 973,391 (0.51\%)   & 19,185,992(3.00 \%)                \\ 
\hline
Total &381,008 & 192,330,821 & 639,517,281  \\ \hline
\end{tabular}
}
\end{table}
\section{Measurement of Security Issues}
\label{sec:security}

In this section, we study the security issues of EOSIO by investigating both the on-chain and off-chain data we collected. We first explore the \textit{permission misuse} phenomenon, and then focus on \textit{real-world security attacks} that we observe in the wild.

\subsection{eosio.code Permission Misuse}
\label{sec:permission}

\subsubsection{Overview of Permission Mechanism.}
EOSIO supports a permission system that can control access to contracts (Section~\ref{subsec:account}).
Users can modify their account permission group and link each permission to different kinds of action. For the default setting, only the \textit{owner} key can sign the \textit{updateauth} action to change the private key of an account, yet custom permissions are incredibly flexible and address numerous possible use cases when implemented. 
Most notably, the \textit{eosio.code} permission is designed to enable contracts to execute inline actions.
However, once a user grants this permission to a contract, it can call system contracts in the name of the user. 
This means that \textit{the eosio.token permission can be assigned to transfer EOS tokens without notifying the user}.
For example, this type of permission misuse has been discovered in a popular DApp named \textit{EOS Fomo3D} (whose account name is \textit{eosfoiowolfs}), which caused serious security threats that many users stop using it~\cite{wolfpermission}.

\begin{figure}[h!]
    \centering
    \subfloat[permission assignment]{
        \label{subfig:permission1}
        \includegraphics[width=0.20\textwidth]{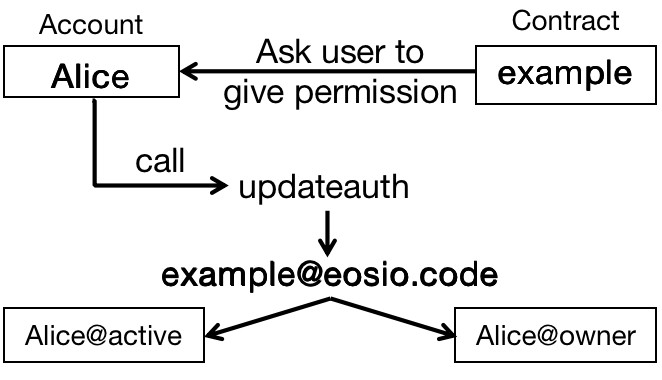}
     }
     \subfloat[permission misuse]{
        \label{subfig:permission2}
        \includegraphics[width=0.255\textwidth]{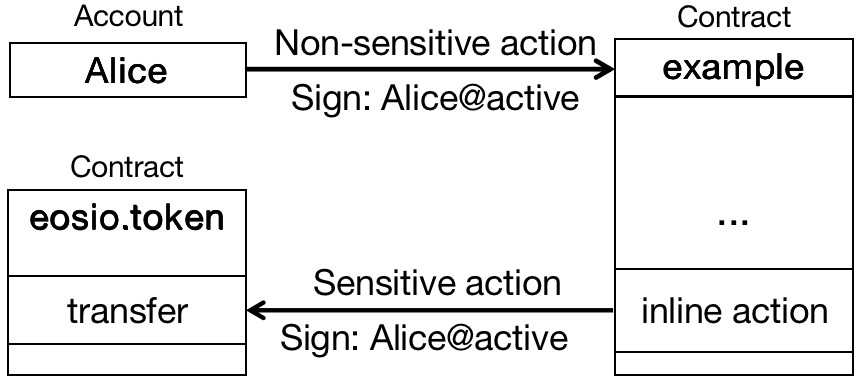}    
     }
     \caption{Illustration of permission misuse}
     \label{fig:permission_misuse}
\end{figure}

\subsubsection{Overview of eosio.code Permission Misuse}
Fig.~\ref{fig:permission_misuse} illustrates how \textit{eosio.code} permission misuse occurs. 
First, a smart contract declares that the \textit{eosio.code} permission is needed in the code.
A users then grants \textit{eosio.code} during the contract invocation, through another two permissions, i.e., \textit{owner} and \textit{active}. 
Specifically, users first link the \textit{eosio.code} of one contract to the \textit{owner}/\textit{active} permission using updateauth action (Figure~\ref{subfig:permission1}).
By doing so, for every action signed by \textit{owner}/\textit{active}, the smart contract can invoke inline actions in the name of user (Figure~\ref{subfig:permission2}).

The above is is typically used for convenience, \eg allowing a game to automatically initiate transfers. However, in practice, the permission opens up a number of attacks. 
The community suggests that contract developers should get rid of eosio.code~\cite{PermissionGrant, securecoding}, and recommends users to reserve their eosio.code permissions unless they trust that contract.

\subsubsection{Detecting eosio.code Permission Misuse}
As permission changes are related to action \textit{updateauth}, we can track on-chain data to see whether there exists any permission misuses. 
Once we find an account grants the \textit{eosio.code} permission to smart contract accounts with different public keys\footnote{Note that we compare the public keys of the involved accounts in order to filter the \textit{valid} permission grants for users with multiple accounts.}, we identify this behavior as a \textit{potential permission misuse}. 
As aforementioned~\ref{subsec:account}, there exists a threshold that must be reached to authorize the execution of the action.
Thus, only when the \textit{assigned permission weight} surpasses the \textit{permission threshold} can another account invoke the corresponding permission. Otherwise, it may require several accounts to authorize a permission invocation together. 
As a result, for each permission grant, we further analyze the weight of the accounts who have been granted the \textit{eosio.code}, and then compare it with the user-defined threshold, to identify the \textit{permission misuse} finally. 

\begin{figure}[h]
\small
  \centering
  \includegraphics[width=0.85\linewidth]{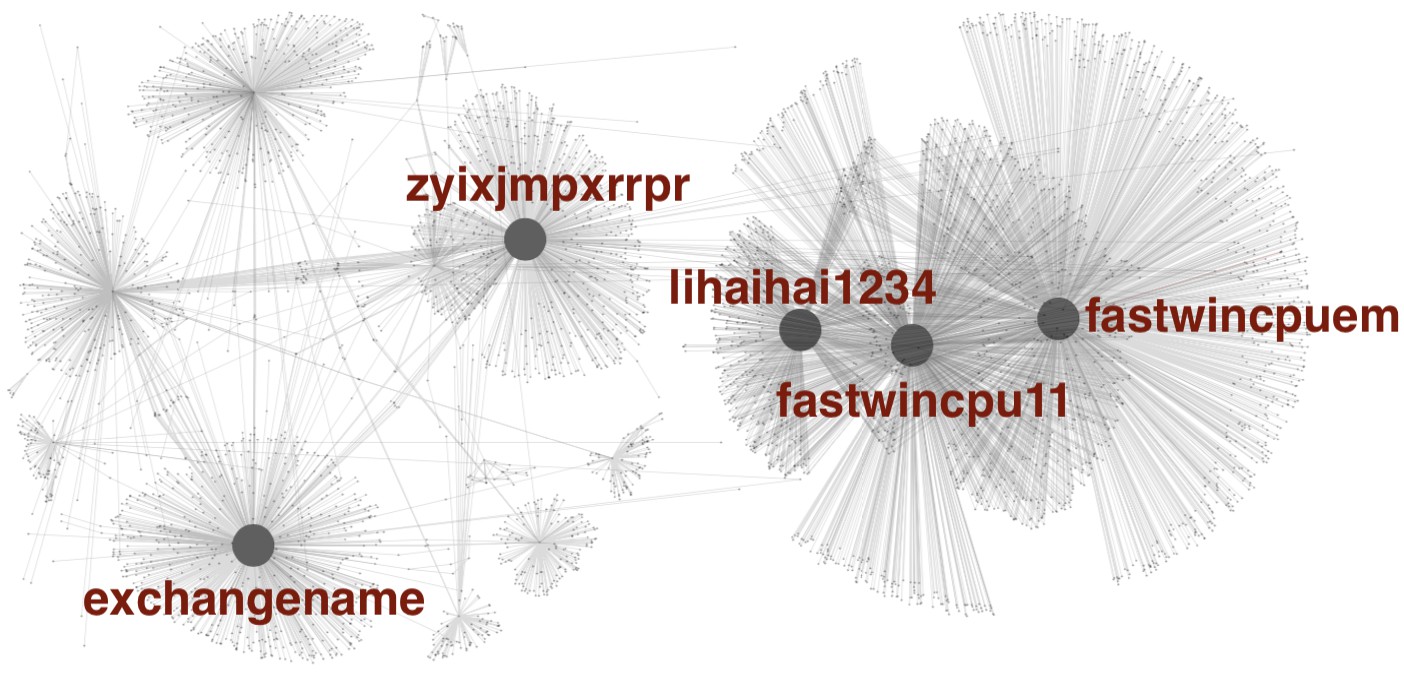}
  \caption{Visualization of the permission misuse.}
  \label{fig:permission_graph}
\end{figure}

\subsubsection{Results.}
In total, we have collected $327,287$ \textit{updateauth} actions. Among them, $26,899$ are associated with the permission \textit{eosio.code}. After filtering out actions whose granted weight is lower than the threshold, we identify $18,321$ permission grant actions in total, and $72.25\%$ of them are linked to \textit{active} permission.
Thus, we have identified 13,237 permission misuse in total, involving 5,541 user accounts, who grant their permissions to 407 smart contracts. The top 5 contracts that have been granted \textit{eosio.code} with \textit{active} are shown in Table~\ref{table:permission_misuse}.
Figure~\ref{fig:permission_graph} illustrates the permission misuse graph, with each node denotes an account, and each edge represents a permission grant action. We observe that most of the smart contracts that require sensitive permissions are gambling games (i.e., 4 in Top-5). Contract \textit{fastwincpuem} is the most representative one.
To play the game, over 1,400 users have to grant the permission to it, which pose serious security risks.

\begin{table}[tb!]
\centering
\caption{Top accounts with granted eosio.code permission.}
\label{table:permission_misuse}
\resizebox{\linewidth}{!}{
\begin{tabular}{l|l|l|l}
\hline
\textbf{linked permission} &\textbf{Account} & \textbf{\# auth account} & \textbf{Identity} \\ \hline \hline

\multirow{5}{*}{active} & fastwincpuem & $1,453$ & gambling game \\
                    & fastwincpu11 & $1,231$ & gambling game \\
                    & zyixjmpxrrpr & $672$ & gambling game \\
                    & exchangename & $626$ & Account exchange \\
                    & lihaihai1234 & $494$ & gambling game \\
\hline
 \end{tabular}
}
\end{table}

\subsection{Real-World Attacks}

Although a number of security reports~\cite{eos-attack1, eos-attack2, eos-attack3} have revealed attacks, it is still unknown how prevalent they are. 
By surveying \textbf{37} attack reports~\cite{peckshield,slowmist}, we next manually classify attacks into two main categories: \textit{Unchecked Input Attacks} and \textit{Predictable State Attacks}. 
Using these observations, we have implemented a monitoring system to identify suspicious actions and accounts.

\subsubsection{Unchecked Input Attacks} 
\label{subsec:unchecked_inputs}
We identify 4 samples of attacks that exploit unchecked inputs~\cite{peckshield-eos-tutorial}.
These rely on victim contracts failing to properly validate the identities of received tokens or notifications~\cite{peckshield-eos-tutorial}. 
It has two variants:
    (1) \textbf{Fake EOS Transfer Attack}, which transfer fake EOS tokens to deceive the victim contract into believing that it is receiving EOS tokens~\cite{fake-eos-transfer}.
    (2) \textbf{Fake EOS Notice Attack}, which sends fake notifications to deceive the victim contract into believing that it is receiving EOS tokens~\cite{fake-eos-notice}.
    
We devise a pattern-based mechanism to detect these two types of attacks, by looking for the patterns summarized in Table~\ref{table:patterns_uncheckedinput}~\cite{fake-eos-transfer,fake-eos-notice}. {\cmark} means the pattern is applicable to detect that particular attack, whereas {\xmark} indicates that the pattern is not applicable. In addition, pattern with \textit{Yes} means the condition must be met; alternatively, pattern with \textit{No} has to be otherwise satisfied.
In total we have 2 patterns for the \textit{Fake EOS Transfer Attack} and 3 patterns for the \textit{Fake EOS Notice Attack} respectively.
Note that the aforementioned patterns may lead to false positives. For instance, the transfer token symbol can always be specified as ``EOS'' by the sender for any non-malicious (e.g., testing) purpose.
Thus, we further use heuristic strategies to alleviate this issue. Specifically, for the \textit{fake EOS transfer attack}, if an account \textit{A} sends fake EOS tokens to account \textit{B}, and on the same day account A makes profit from B, we then mark A as suspicious. Similarly, as to the \textit{fake EOS notice attack}, we further verify all accounts that have sent fake EOS notice to a DApp's account. These heuristics can identify all 4 unchecked input attacks.

\begin{table}[t!]
\centering
\caption{Patterns of Fake EOS Transfer/Notice Attacks.}
\label{table:patterns_uncheckedinput}
\resizebox{\linewidth}{!}{
\begin{tabular}{c|c|c}
\hline
\textbf{Pattern}& \textbf{Fake EOS Transfer} & \textbf{Fake EOS Notice} \\
\hline
\hline
\tabincell{c}{The transfer token symbol is EOS.}    & \cmark & \cmark \\
\hline
\tabincell{c}{Whether the contract executes ``transfer'' action\\ is the official contract (eosio.token) or not.}       & No  & Yes \\ 
\hline
\tabincell{c}{The notice receiver is the victim, rather\\ than any accounts involved in the transfer.}    & \xmark & \cmark \\ 
\hline
\end{tabular}
}
\end{table}

\subsubsection{Predictable State Attacks}
\label{subsec:predictable_state}
We collect 33 samples of attacks that rely on predictable state. These include variations such as: 
(1) The \textbf{Random Number Attack}, which exploits the vulnerabilities present in the procedure to generate random numbers. (2) The \textbf{Roll Back Attack}, which defrauds the lottery game without actually paying the bet cost by rolling back the corresponding unsatisfied reversible transaction~\cite{rollback-blacklist,rollback-replay}. 
(3) The \textbf{Transaction Congestion}, which launches a DoS attack by sending a large number of deferred transactions based on the predictable state, to force the victim contract to regenerate a different result~\cite{congestion}.
This attack is usually targeted at gambling games (32 out of 33 attack gambling games).

Such attacks cannot be identified using the heuristics in Section~\ref{subsec:unchecked_inputs}. Hence, we turn to another intuition: \textit{accounts with large profits in a short period of time with high frequency may be suspicious}. 
This assumption, however, could lead to many false positives. To investigate this, we manually analyze the collected attack accounts and summarize several characteristics:

\begin{enumerate}
\item \emph{Although the profit attackers earn varies widely, their profitability ratio\footnote{Here, we define the ratio as the amount of EOS received over the amount of EOS sent.} is relatively high.} In contrast, even though a normal account has the potential to earn high profits (e.g., if its wins a large bet), it is unlikely that normal accounts consistently maintain a high profitability ratio. For the collected attacks, the median profit is $2,202$ EOS, and the median of profitability ratio is $2.4$.

\item \emph{Attacks usually happen over a relatively short time} (e.g., from a few minutes to an hour). This is quite different to other normal behaviors. As a reference, all the $33$ Predictable State Attacks finish within an hour.

\item \emph{The total amount of the ``excessive'' profit always makes-up a large portion of the total volume of the transactions for the account with the target DApp}. Because attack accounts are usually active for just a few days. Once attackers succeed, they usually do not use the Dapp anymore. In contrast, normal users with high profits have no reasons to be ``silent''.
\end{enumerate}

Based on these observations, we are able to propose a semi-automated approach to label suspicious accounts, as follows: 

\textbf{Step 1.} We first monitor the accounts (in the EMFG) that gain high profit with a high profitability ratio. For a given account, if the profit is larger than a threshold \textit{W1} and the profitability ratio is larger than a threshold \textit{W2} in some granularity of time period, we mark it as suspicious. In particular, we specify two kinds of granularity: \textit{one day} for the coarse-grained detection and \textit{an hour} for the fine-grained detection, to achieve a good accuracy with acceptable performance. 
Note that, we empirically set $W1=400$, $W2=1.2$. This is a looser boundary (compared with known attacks) to identify as many suspicious accounts as possible.
This step is capable of filtering out most of the unrelated normal accounts.

\textbf{Step 2.} For a given flagged victim DApp candidate, we record the profits and the corresponding dates for all suspicious accounts. For each suspicious account, we further calculate the proportion of profits (denoted by $p$) over the total incoming EOS tokens received from that DApp. An account will be labelled as highly suspicious and require further verification if $p$ is larger than a threshold \textit{W3}, which gives the liveness of one account to some extent. We empirically set $W3=0.9$, i.e., the excessive profits must have occupied at least $90\%$ of all the money transfers. As a reference, for the collected attack accounts, all of them have achieved $100\%$ on this metric.

\textbf{Step 3.} 
The aforementioned \emph{profit-driven} approaches are able to identify most of the attacks, as the sole purpose of the attacks we considered is to make profit. However, it is quite possible that we may include some benign behaviors, due to the limitation of the thresholds we defined. 
Thus, to filter out possible false positives, our last step is to manually analyze the remaining suspicious attack accounts and their behaviors. Note that, previous steps have filtered most of the accounts and actions, making our manually analysis possible.
\emph{Our goal is to either replay the ``attacks'' they performed, or find more evidences, to confirm whether a suspicious account fulfills the characteristics of the attacks.}. 
To the best of our knowledge, this is the only reliable way to label attacks.
We employ three techniques.
(1) Random number attacks can be reproduced on the testnet we built.
Thus, for each suspicious account, we first analyze whether it targets a gambling Dapp. Then we repeat its actions to see whether we are able to launch random number attacks.
We have implemented Proof-of-Concept (PoC) scripts on our testnet to verify the random number attacks are indeed performed.
Our implementation follows some attack scripts provided by the community~\cite{PoCRandomNumber}.
(2) For the roll back attack, we take advantage of the EOSIO client we customized to synchronize with the mainnet. Although roll back actions will not be put on the chain, they will be broadcast and a client could receive them. For each account, we further analyze their broadcast information. If one account has too many roll back transactions with a high profit, we believe it launches roll back attacks.
(3) The transaction congestion attack is quite noticeable, as attackers usually create thousands of deferred transactions and may even paralyze the whole mainnet. Thus, for each account, we analyze its deferred transactions to see whether it performs DDoS-like attack.
In brief, the manual investigation allows us to determine the attacks, the corresponding attack accounts, and the victim DApps.

\subsubsection{Detection Result}

Using the above detection mechanisms, we identify the presence of attacks across the entire EOSIO blockchain. We discover $301$ attack accounts associating with $1,518,401$ EOS tokens (roughly 4.8 million US\$). Table~\ref{table:top-attack} lists the top-3 of them.
Specifically, we discover $24$ attack accounts associated with $136,881$ EOS tokens for the \textit{fake EOS transfer attack}s, and $28$ attack accounts associating with $235,867$ EOS tokens for the \textit{fake EOS notification attacks} respectively.\footnote{Two accounts are related to both fake EOS transfer and fake EOS notification attacks.}
Furthermore, we found $251$ accounts ($1,070,005$ EOS tokens) that rely on predictable state.

These attacks are targeting 112 victim DApps. Note that, we report all the identified attacks to the corresponding DApp teams. 
By the time of our study, $80$ attack accounts have been confirmed by them, causing $828,824$ EOS tokens in losses (roughly 2.6 million US\$).
For the remaining attack accounts being detected, We are still working with DApp developers to make final confirmation.\footnote{Some DApp teams are no longer active due to the financial loss of attacks.} 
Apart from notifying the DApp developers for a timely damage control, we also assist them in tracing the losses and provide a digital forensic service to support advanced collaboration with other third parties like exchanges. For example, we have assisted the development teams of DApp BetDice, ToBet, EOS MAX, BigGame in \textbf{tracing the losses of 288,329 EOS} by providing them attack evidence.

\begin{table}[tb!]
\centering
\caption{Top-3 attack accounts identified.}
\label{table:top-attack}
\resizebox{\linewidth}{!}{
\begin{tabular}{r|l|l|l}
\hline
\textbf{Attackers}& \textbf{target} & \textbf{profit} &\textbf{confirmed}\\
\hline
\hline
\hspace{0.5cm}
hnihpyadbunv   & Dapp Dice &  $195,530$ & Yes \\
ilovedice123    & EOSBet  & $137,970$ & Yes \\ 
eykkxszdrnnc    & EOS Max, BigGame & $71,731$ & Yes \\ 
\hline
\end{tabular}
}
\end{table}
\section{Limitations and Implications}

\textbf{Limitations.} 
To the best of our knowledge, this is the first comprehensive study of EOSIO. However, our study carries some limitations. 
In several cases, we have relied on heuristics and manual validation. 
This was necessary due to the paucity of ground-truth data. For example, for the bot-like account detection, we have tried our best to validate our approach, however, it is impossible for us to make sure \emph{all} accounts flagged are truly bots. 
Similarly, although we only focus on bot-like accounts that operate in groups and have similar behaviors, it is possible that there are bots that do not fall into this definition.
We raise similar observations regarding our attack detection. We have applied heuristics to flag accounts that are suspicious, and relied on manual efforts to confirm them.
This, of course, might not be scalable and could mean that we only offer a lower-bound. Future work will involve incorporating program analysis and dynamic testing techniques to help us automatically identify attacks.

\noindent \textbf{Implications.}
Our observations are of key importance to stakeholders in the community.
First, considering the large number of fraudulent behaviors and security issues we discover, the governance of EOSIO needs to be improved. 
Second, we argue DApp developers should take actions immediately to address the security issues introduced by vulnerabilities of smart contracts.
Third, as mentioned earlier, we provide a digital forensic service to help developers recoup the losses by collaborating with other third parties like exchanges. As of this writing, we have helped four DApp teams in tracing the losses by providing them attack evidence. 
Moreover, our findings could help users and investigators to understand the status quo of EOSIO ecosystem, and protect them from being deceived by some ``popular'' DApps.
Last but not least, our findings and techniques could be generalized to other blockchain platforms as well. For example, it is reported that blockchain bots were also found in the TRON DApp ecosystem~\cite{TronBots}. Our observations and techniques could be easily adopted to detect them.
\section{Related Work}
\label{sec:related}

\textbf{Transaction-based Analysis.}
Previous researchers have investigated blockchain systems by performing transaction-based analyses. Several works focus on Bitcoin~\cite{FergalSC2011, DoritFC13, alex2014deanonymisation, ChenDF15, MichaelArxiv15, DamianoCN16, SilivanxayICDMW18}, including de-anonymization and money laundering detection, by using graph-based approaches. Researchers have also investigated Ethereum by using transaction-based analyses~\cite{WrenICITST17, TingINFOCOM18}.
For example, Chen et al.~\cite{TingINFOCOM18} gives a graph-based analysis of Ethereum, which is somewhat similar to our approach in Section~\ref{sec:general}. However, our work differs from previous studies because of the differences between EOSIO and other platforms. More importantly, we have proposed systematic approaches to successfully identify and measure bot-like accounts and malicious accounts that have never been identified before. To the best of our knowledge, this is the first comprehensive work to study behaviors on EOSIO in a systematic manner.

\noindent \textbf{Vulnerability/Attack Detection.}
There are a number of studies focus on analyzing the vulnerabilities in Ethereum smart contracts~\cite{LoiCCS16, ChristofACSAC18, liu2018reguard, tikhomirov2018smartcheck, jiang2018contractfuzzer, tsankov2018securify, SukritNDSS18}.
For example, Luu \textit{et al.}~\cite{LoiCCS16} built a symbolic execution tool to discover potential vulnerabilities. Kalra \textit{et al.}~\cite{SukritNDSS18} provided a static analysis tool to detect vulnerable smart contracts based on LLVM. He \textit{et al.}~\cite{he2019characterizing} characterized code-clone based vulnerabilities in Ethereum smart contracts.
Regarding EOSIO, however, most are reports~\cite{random-number, fake-eos-transfer, fake-eos-notice, rollback-blacklist, rollback-replay, congestion} focusing on attack/vulnerability analyses by security companies; only a few academic publications~\cite{SangsupWOOT19, EVulHunter} are available. Lee \textit{et al.}~\cite{SangsupWOOT19} introduced and studied four attacks stemming from the unique design of EOSIO. Quan \textit{et al.}~\cite{EVulHunter} focused on detecting two types of vulnerability for EOSIO smart contracts. A number of previous works have proposed tools to detect vulnerabilities by performing static analyses. In contrast, we adopt a statistical approach to detect attacks by analyzing the anomalies. 

\section{Conclusion}

In this paper, we have performed the first large-scale measurement study of the EOSIO blockchain. By constructing a comprehensive dataset, we first analyzed the activities including money transfer, account creation and contract invocation. We further focused on security and fraudulent issues, including bot-like accounts and attack detection.
Our exploration has identified many security issues and revealed various interesting observations, including thousands of bot accounts, hundreds of real-world attacks, as well as insights for future research directions.

\bibliographystyle{ACM-Reference-Format}
\balance
\bibliography{main}

\end{document}